\documentclass[aps,prd,twocolumn,superscriptaddress,nofootinbib]{revtex4-1}

\usepackage{amssymb}
\usepackage{amsmath}
\usepackage{graphicx}
\usepackage{enumerate}
\usepackage{mathtools}
\usepackage{color}
\usepackage{bbold}
\usepackage{subfigure}
\usepackage{slashed}
\usepackage{xcolor}
\usepackage{bm}

\begin{document}


\title{Fast oscillations, collisionless relaxation, and spurious evolution of supernova neutrino flavor}


\author{Lucas Johns}
\email[]{NASA Einstein Fellow \\ ljohns@berkeley.edu}
\affiliation{Department of Physics, University of California, Berkeley, CA 94720, USA}

\author{Hiroki Nagakura}
\affiliation{Department of Astrophysical Sciences, Princeton University, 4 Ivy Lane, Princeton, NJ 08544, USA}

\author{George M. Fuller}
\affiliation{Department of Physics, University of California, San Diego, La Jolla, CA 92093, USA}

\author{Adam Burrows}
\affiliation{Department of Astrophysical Sciences, Princeton University, 4 Ivy Lane, Princeton, NJ 08544, USA}

\begin{abstract}
Mounting evidence indicates that neutrinos likely undergo fast flavor conversion (FFC) in at least some core-collapse supernovae. Outcomes of FFC, however, remain highly uncertain. Here we study the cascade of flavor-field power from large angular scales in momentum space down to small ones, showing that FFC enhances this process and thereby hastens relaxation. Cascade also poses a computational challenge, which is present even if the flavor field is stable: When power reaches the smallest angular scale of the calculation, error from truncating the angular-moment expansion propagates back to larger scales, to disastrous effect on the overall evolution. Essentially the same issue has prompted extensive work in the context of plasma kinetics. This link suggests new approaches to averting spurious evolution, a problem that presently puts severe limitations on the feasibility of realistic oscillation calculations.
\end{abstract}

\maketitle

\section{Introduction}

Recent advances suggest that neutrino flavor transformation in core-collapse supernovae may occur on much shorter scales than previously believed. This new class of collective effects has been dubbed fast flavor conversion (FFC), in reference to the fact that the associated instabilities grow at rates proportional to the neutrino self-coupling potential $\mu = \sqrt{2} G_F n_\nu$ \cite{sawyer2005, sawyer2009, sawyer2016, chakraborty2016b, chakraborty2016c, dasgupta2017, dasgupta2018, dasgupta2018b, airen2018, abbar2018, abbar2019, capozzi2019, capozzi2019b, yi2019, chakraborty2020, martin2020, johns2020, bhattacharyya2020, abbar2020, shalgar2020b, capozzi2020, xiong2020, shalgar2020, bhattacharyya2020b}. Although a comprehensive appraisal of their prevalence and importance is still lacking, fast instabilities have now been located in several simulations of core-collapse supernovae and accretion-disk systems in remnants of coalescing neutron-star--neutron-star and neutron-star--black-hole binaries \cite{tamborra2017, wu2017, wu2017b, abbar2019b, azari2019, shalgar2019, nagakura2019, morinaga2020, glas2020, abbar2020b, padilla2020, george2020}. It appears quite possible that FFC engenders significant dynamical and compositional changes in these environments.

In Ref.~\cite{johns2020} we proposed that FFC stems from specific features of the equations of motion that become more apparent upon expanding in angular moments. Taking the neutrino density to be high, the neutrino angular distributions to be axially symmetric, and the matter background to be homogeneous on the scales of interest, neutrino flavor evolves on short timescales according to the equation
\begin{equation}
\dot{\mathbf{P}}_v + v \mathbf{P}'_v \cong \mu \left( \mathbf{D}_0 - v \mathbf{D}_1 \right) \times \mathbf{P}_v, \label{pveom}
\end{equation}
where $\mathbf{P}_v$ is the polarization vector of neutrinos with propagation angle $v = \cos\theta$ ($P_{v,z}$ being its flavor content), $\mathbf{D}_l$ is the $l$th Legendre moment of $\mathbf{P}_v - \mathbf{\bar{P}}_v$, and the overdot and prime denote derivatives with respect to time and space. In the limit that the flavor field is homogeneous, FFC is the result of pendulum-like motion of $\mathbf{D}_1$, a point supported by linear stability and numerical analysis. In the other analytically tractable limit, where the flavor field is stationary, FFC appears to be the result of pendulum-like motion of $\mathbf{D}_0$. Although the conjecture regarding steady-state solutions is supported by direct manipulation of the equations of motion, we have not attempted to test it with stability analysis or numerics.

In the numerical calculations of Ref.~\cite{johns2020}, FFC is observed to be nearly periodic. Aperiodicity occurs because each angular moment is coupled to its neighbors in an infinite chain, and nothing prevents power from making its way out to ever higher $l$, never to return. On longer time scales the influence of finite $\omega = \delta m^2 / 2p$ is felt as well, $p$ being neutrino momentum and $\delta m^2$ the mass-squared difference. Crucial conservation laws are broken by $\omega \neq 0$, and slow collective effects (with growth rates proportional to $\sqrt{\omega \mu}$) come into play, interacting with fast oscillations.

Whereas in our previous study we emphasized the importance of low-$l$ moments in shaping FFC, here we focus on the dynamics at later times and smaller angular scales. Central concepts include irreversibility and collisionless relaxation. Under unitary evolution, there is only one way for a dense system of neutrinos to relax to flavor equilibrium: by transferring power to smaller scales in phase space. The asymptotic state is not a true equilibrium, which can only be achieved with collisional processes, but it is equilibrated in a coarse-grained sense, and for practical purposes the coarse-grained flavor field is what matters most. Viewing neutrino flavor evolution through the lens of collisionless relaxation brings out more clearly some of its important parallels with other physical systems. We discuss phase-space transfer at a general level in Sec.~\ref{analogues}.

Flavor relaxation via phase-space transfer has been observed and discussed in numerous references, often with the appellations \textit{kinematic decoherence} and \textit{multi-angle decoherence}, to be contrasted with the collisional variety. Coarse-grained equilibration has been witnessed repeatedly in numerical toy models. Yet the theory describing how the flavor field relaxes---on what time scale, to what asymptotic ensemble, and with what parametric dependence---is underdeveloped. In our view, this is one of the most important gaps in our current understanding of the supernova oscillation problem, as it is the missing link connecting linear instabilities and idealized collective phenomena to realistic neutrino transport. Had we powerful enough computers, the full problem could be simulated and the physics we seek to understand would be incorporated organically. But alas we do not, and given the astrophysical stakes, waiting for computing power to catch up does not strike us as prudent.

The foundational exposition of multi-angle decoherence was given by Raffelt and Sigl in Ref.~\cite{raffelt2007b}, where they showed that a nearly isotropic gas of neutrinos is prone to rapid flavor depolarization. One of the important achievements of that paper was to trace the origin of depolarization to exponential growth of $\mathbf{D}_1$ on a time scale proportional to $\sqrt{\omega\mu}^{-1}$. Recent years have seen this phenomenon reinterpreted as an example of a multi-zenith-angle (MZA) instability. But while linearized dispersion relations are a powerful analytic tool, they do not supersede hard-won insights at the level of the polarization vectors. The goal, after all, is not just to locate instabilities, but to understand their consequences.

Shortly after the publication of Ref.~\cite{raffelt2007b}, it was established that realistic asymmetries between the fluxes of $\nu_e$ and $\bar{\nu}_e$ typically prevent this instability from taking hold during deleptonization \cite{esteban2007} (with possible caveats now coming from the emerging understanding of \textit{global} asymmetries in the supernova neutrino emission \cite{vartanyan2019}). Still, the observations regarding multi-angle decoherence remain interesting, especially in light of the connection between FFC and pendulum-like motion of $\mathbf{D}_1$. The exponential-growth solution is accompanied by rapid decoherence because, in accord with a conservation law in the equations of motion [Eq.~\eqref{esenergy} below], the growth occurs at the expense of $\mathbf{S}_0$, the $0$th Legendre moment of $\mathbf{P}_v + \mathbf{\bar{P}}_v$. In homogeneous FFC, however, the vector magnitude $D_1$ is constant. If \textit{fast} instability is to be complicit in multi-angle decoherence, it must accomplish this end differently than the slow instability identified in Ref.~\cite{raffelt2007b}.

We show in Sec.~\ref{cascade} that FFC hastens relaxation by a distinct mechanism in which the flavor field is kinematically decohered by polarization-vector dephasing.\footnote{Here and in the following we use ``dephasing'' to refer to nonzero $\mathbf{A}\times\mathbf{B}$ for two polarization vectors $\mathbf{A}$ and $\mathbf{B}$. Most often, the relevant dephasing will be with respect to $\mathbf{D}_1$. We use ``kinematic decoherence'' to encompass any kind of collisionless depolarization.} This mechanism, whereby power leaks to smaller scales in momentum space, is in fact always operative, even when the flavor field is linearly stable (see below). Any amount of anisotropy is enough to expose the system to this source of effective irreversibility.

Regardless of its origin, momentum-space transfer is a rather grave computational concern. Once power cascades down to the smallest angular scale of the calculation, the flavor field begins to feel the finite resolution of momentum space. Errors at the smallest scale become magnified as they propagate back up the chain of multipoles. As we show in Sec.~\ref{numerical}, even a minuscule amount of power slipping out to high $l$ is enough to wreak havoc.

One solution, of course, is to evolve a large number of multipoles, but since the details at very high $l$ are practically irrelevant, this is a highly wasteful approach in a setting where one can ill afford to be profligate. A better solution would be to close the equations at a reasonable $l$ in such a way as to avoid spurious backreaction. Consistent with our previous study, we find that \textit{fast oscillations} are decently well approximated by evolving only the largest angular scales. But for \textit{relaxation} to occur, the low-$l$ moments must be able to lose power to the ``inertial range'' that forms at higher $l$. In this sense, cascade is a necessary evil. We do not solve the closure problem here, but at the end of Sec.~\ref{numerical} we indicate possible paths forward and point to some of the strategies that have been employed in plasma kinetics.

In Sec.~\ref{conc} we summarize our analysis. The results presented here are based on a simplified model of flavor evolution and are by no means direct predictions of real-world outcomes. The issues we address, however, are quite general. If it turns out that instabilities in realistic settings prompt rapid flavor equipartition (as assumed in Refs.~\cite{wu2017b}, for example), or something comparably simple and robust, then the details of how the flavor field relaxes might be purely academic. But failing that, the physics of relaxation must be understood more deeply, and momentum-space transfer will be an essential element. We speculate, in the final section, on how our findings are likely to be modified in more realistic models.

\section{Collisionless relaxation \label{analogues}}

The transfer of energy across scales in phase space is a nearly universal feature of weakly collisional kinetic systems. In a classical context, it is related most basically to the law of inertia and the Hamiltonian preservation of phase-space density. The latter constrains what collisionless trajectories can accomplish: The best they can do, as far as transforming the distribution in phase space, is to stretch, squeeze, and knead it. Because phase-space transfer, which describes such reconfigurations, tends to move the system toward more typical macroscopic states, it serves as the classical mechanism of collisionless relaxation.

Cascade---by which we mean sustained, directed transfer---occurs when a system's approach to equilibrium is facilitated by preferentially shifting energy to smaller scales than the one at which it is sourced (or, in the case of inverse cascade, to larger ones). Fluid and magnetohydrodynamic turbulence may be the most famous examples: In the classic turbulent system, energy at the driving scale cascades down to smaller scales until viscous dissipation becomes efficient. Fluids, of course, are not weakly collisional, but they are the exception that proves the rule of transfer's commonness. Even under the non-Hamiltonian, momentum-isotropizing effects of collisions, turbulent dynamics takes advantage of the other half of phase space and works across spatial scales. 

Despite neutrino oscillations being a quantum phenomenon, classical kinetics is a good starting point for thinking about weakly collisional neutrino transport. In the quantum-kinetic description \cite{sigl1993, vlasenko2014} (from which Eq.~\eqref{pveom}, for example, derives), physically relevant variations in time and space occur over scales large enough that the Wigner \textit{quasiprobability} distribution resembles a bona fide \textit{probability} distribution. Many-body correlations are also argued, one way or another, to be ignorable \cite{friedland2003, volpe2013}. Neutrino quantum kinetics then appears to describe a theory of individual particles, each carrying quantum degrees of freedom but free-streaming classically. These considerations imply significant resemblances between neutrino flavor evolution and classical kinetics, despite the quantum character of the former.

Direct cascade in a kinetic system corresponds, in coordinate space, to the formation of small-scale inhomogeneities and, in momentum space, to the formation of small-scale spectral distortions and angular features. Coordinate- and momentum-space transfer (of which cascade is a special case) are both already possible in the collisionless, field-less Vlasov equation in one spatial dimension:
\begin{equation}
\dot{f} (t,x,u) + u f' (t,x,u) = 0, \label{simplevlasov}
\end{equation} 
where $f$ is a particle distribution function and $u$ is the particle velocity. If the spectrum is monochromatic, $f$ is simply translated along $x$. But with a spectrum of velocities, inhomogeneity and polychromaticity work together to make for more complicated evolution, involving mode transfer in $x$- and $u$-space. The point, to rephrase the opening remarks of this section, is that particle free-streaming on its own is sufficient to generate smaller-scale features over time. Nonzero fields can enrich the dynamics further, but either way, Liouville's theorem implies that a collisionless Vlasov system can only approach equilibrium in a coarse-grained sense, by hiding the fluctuations below the resolved granularity.

Cascade commonly poses numerical challenges. In high-Reynolds-number fluids, like those found in the convective regions of supernovae, spatially resolving the evolution down to the dissipation scale is often impossible \cite{radice2018}. In a collisionless plasma, phase-space filamentation causes numerical solutions of the Vlasov equation to fail in finite time, an issue that has been the subject of computationally oriented investigation going back decades \cite{joyce1971, klimas1987}. In collisionless gravitational systems, closely related issues arise when solving for the evolution in phase space \cite{rasio1989}. Cascade must be handled delicately in all of these settings. Besides being the cause of numerical trouble, it is also of \textit{physical} importance, being crucial to (among other things) Reynolds stresses in fluids, phase-space structures and enhanced collisionality in plasmas \cite{dupree1972, pezzi2016}, and phase mixing and violent relaxation in gravitational systems \cite{lynden1967}.

In a collisionless gas of neutrinos, there are multiple potential sources and forms of phase-space transfer. The phenomenon most like the previously mentioned examples is the formation of spectral distortions and filamentation due to classical transport:
\begin{equation}
\dot{f}_{\nu_\alpha} (t, \mathbf{x}, \mathbf{\hat{u}}, E) + \mathbf{\hat{u}} \cdot \boldsymbol{\nabla} f_{\nu_\alpha} (t, \mathbf{x}, \mathbf{\hat{u}}, E) = 0, \label{classical}
\end{equation}
where $\alpha$ identifies flavor. Because ultrarelativistic neutrinos all travel with velocity $u \approx c = 1$ (hence the appearance of the unit vector $\mathbf{\hat{u}}$), Eq.~\eqref{classical} lacks the dispersive shearing exhibited by Eq.~\eqref{simplevlasov}. Moving along a streamline, spectral distortions therefore only appear in the angle-integrated distributions. Other than that, Eq.~\eqref{classical} permits all of the same relaxation processes one expects from the collisionless, field-less Vlasov equation.

In practice, however, collisionless relaxation is not a particularly helpful lens through which to view the \textit{classical} transport of supernova neutrinos. It adds little to the usual picture of neutrinos as transiting from an opaque region to a transparent one as they propagate outward in radius. The angular distributions of momentum, for instance, are isotropized by emission, absorption, and scattering, and are rendered more forward-peaked by free-streaming out from a roughly spherical geometry. The microphysics and the environment are simply not conducive to small-scale angular features beyond those already captured by interpolating between the isotropic and single-beam limits. This fact underpins the adequacy of moment methods in neutrino radiative transfer.

The situation is quite different with oscillating neutrinos. Flavor ceases to be conserved along trajectories, and the nonlinearity arising from neutrino--neutrino forward scattering supports collective effects that promote collisionless relaxation. Rather in contrast to collisional processes, oscillations \textit{favor} the generation of small-scale features in the neutrino flavor field of a supernova.

The oscillation terms engineer this outcome in different ways. Ignoring matter currents (and suppressing independent variables, which are the same as above), the equation of motion of polarization vector $\mathbf{P}$ with unit velocity vector $\mathbf{\hat{u}}$ is
\begin{align}
\dot{\mathbf{P}} + \mathbf{\hat{u}} \cdot \boldsymbol{\nabla} \mathbf{P} = \bigg[ \omega &\mathbf{B} + \lambda \mathbf{L} + \sqrt{2} G_F \int d\Gamma' \mathbf{D}' \notag \\
&- \sqrt{2} G_F \int d\Gamma' \left( \mathbf{\hat{u}}\cdot\mathbf{\hat{u}}' \right) \mathbf{D}' \bigg] \times \mathbf{P}. \label{quantum}
\end{align}
Here, $d\Gamma'$ is a phase-space element, with the integrals over all (anti)neutrino momenta at $(t, \mathbf{x})$; $\lambda = \sqrt{2} G_F n_e$ is the matter potential; and $\mathbf{B}$, $\mathbf{L}$, and $\mathbf{D}$ are the mass, flavor, and difference vectors, respectively.

The terms on the right-hand side of Eq.~\eqref{quantum} affect phase-space transfer in distinct ways. The dispersive $\omega \mathbf{B}$ term elicits transfer in $p$-space---something that we noted was not possible in Eq.~\eqref{classical}---by causing neutrinos of different energies to dephase. This effect is sometimes (but not always) what has been meant by the term \textit{kinematic decoherence}; it represents the momentum-space complement to wave-packet separation in coordinate space \cite{akhmedov2016}. Like all of the terms, $\omega \mathbf{B}$ interacts with convection. Structures in $p$-space that are generated by it in one spatial region thus get communicated to other regions, and vice versa. Under the common convention that antineutrinos obey the same equations but with $\omega \rightarrow - \omega$, the vacuum term is also responsible for dephasing neutrinos from antineutrinos.

$\lambda \mathbf{L}$ is independent of neutrino momentum and therefore has no \textit{direct} effect on $\mathbf{p}$-space transfer. This is why it can be rotated away in settings that are fully homogeneous, neutrino flavor field included. The term is not entirely without consequence, however, even in the fully homogeneous case, as exemplified by the logarithmic slowdown of the bipolar instability \cite{hannestad2006}. In settings that are inhomogeneous in the flavor field, if not also in the matter background, it is responsible for multi-angle matter suppression \cite{esteban2008}. From one viewpoint, this phenomenon is due to the compression of oscillation patterns along non-radial trajectories. Equivalently, it can be seen as a dephasing effect associated with convection. This is the sense in which $\lambda \mathbf{L}$ alters $\mathbf{p}$-space transfer indirectly.

In the homogeneous limit, which we adopt in the rest of the paper, the isotropic nonlinear term $\int d\Gamma' \mathbf{D}'$ synchronizes neutrinos and antineutrinos of different momenta by acting equally on all of them \cite{samuel1993, kostelecky1995, qian1995a, pastor2002, johns2016, johns2018}. The anisotropic nonlinear term $\int d\Gamma' \left( \mathbf{\hat{p}}\cdot\mathbf{\hat{p}}' \right) \mathbf{D}'$, on the other hand, causes dephasing because of its dependence on $\mathbf{\hat{u}}$. If the system is sufficiently simple or symmetric, (nearly) anisotropic collective oscillations can result; the dephasing need not be permanent. One of the main points of this paper, though, is that realistically this term always drives \textit{some} dephasing that is effectively irreversible. To some extent, if the environment is stationary \textit{and axially symmetric}, the roles of the nonlinear terms are swapped, $\int d\Gamma' \left( \mathbf{\hat{p}}\cdot\mathbf{\hat{p}}' \right) \mathbf{D}'$ acting to synchronize and $\int d\Gamma' \mathbf{D}'$ acting to dephase \cite{johns2020}. While the stationary and homogeneous limits give intuition, in general the flavor field is spatiotemporally evolving, and the influence of axial asymmetry may be significant.

In this paper we focus exclusively on momentum-space transfer, using monochromatic (or integrated-over) spectra for simplicity's sake. Recent progress in the physics of collective neutrino oscillations has brought to the fore the significance of momentum-space anisotropy for flavor \textit{instability}. Our goal is to develop a deeper understanding of how anisotropy enables flavor \textit{relaxation}. As a starting point, and to make contact with Ref.~\cite{johns2020}, we also adopt axial symmetry. We comment on the extensions to axial asymmetry and $\mathbf{x}$-space transfer in the concluding section.

\section{Oscillations and cascade \label{cascade}}

For this analysis, we consider homogeneous, axially symmetric, collisionless flavor evolution, and we assume that the neutrino system is functionally monochromatic due to the high neutrino density. These simplifications allow us to isolate certain fundamental aspects of the momentum-space dynamics. In Sec.~\ref{conc} we offer some comments regarding the effects of inhomogeneity, axial asymmetry, and collisions. Effects associated with a spectrum of energies are not pursued here.

\subsection{Analysis of collective effects \label{analysis}}

With the assumptions stated above, and with the matter potential rotated out of the problem, the equations of motion of the neutrino and antineutrino polarization vectors are
\begin{align}
&\dot{\mathbf{P}}_v = +\omega \mathbf{B} \times \mathbf{P}_v + \mu \left( \mathbf{D}_0 - v \mathbf{D}_1 \right) \times \mathbf{P}_v, \notag \\
&\dot{\bar{\mathbf{P}}}_v = -\omega \mathbf{B} \times \mathbf{\bar{P}}_v + \mu \left( \mathbf{D}_0 - v \mathbf{D}_1 \right) \times \mathbf{\bar{P}}_v. \label{homogeom}
\end{align}
(Relative to Eq.~\eqref{pveom} we have dropped the convective term and restored the vacuum term.) Forming the sum and difference vectors $\mathbf{S}_v = \mathbf{P}_v + \mathbf{\bar{P}}_v$ and $\mathbf{D}_v = \mathbf{P}_v - \mathbf{\bar{P}}_v$, then taking Legendre moments, one has
\begin{align}
&\dot{\mathbf{S}}_l = \omega \mathbf{B} \times \mathbf{D}_l + \mu \mathbf{D}_0 \times \mathbf{S}_l \notag \\
&\hspace{.85 in}- \frac{\mu}{2} \mathbf{D}_1  \times \left( a_l \mathbf{S}_{l-1} + b_l \mathbf{S}_{l+1} \right), \notag \\
&\dot{\mathbf{D}}_l = \omega \mathbf{B} \times \mathbf{S}_l + \mu \mathbf{D}_0 \times \mathbf{D}_l \notag \\
&\hspace{.85 in}- \frac{\mu}{2} \mathbf{D}_1  \times \left( a_l \mathbf{D}_{l-1} + b_l \mathbf{D}_{l+1} \right),
\end{align}
where $a_l = 2l / (2l+1)$ and $b_l = 2 (l+1) / (2l+1)$ \cite{raffelt2007b}.

With $\mu \gg \omega$, $\mathbf{D}_0$ is constant on $\mu^{-1}$ timescales. Fast instability must therefore be caused by instability in $\mathbf{D}_1$. To understand the evolution of $\mathbf{D}_1$ on short timescales, it is helpful to switch to a frame rotating with frequency $\mu D_0$ about $\mathbf{\hat{D}}_0$. We will use primes to denote vectors in the rotating frame. Following Ref.~\cite{johns2020}, we define a (unit-length) ``pendulum vector''
\begin{equation}
\boldsymbol{\delta}' = \frac{\mathbf{D}'_1}{D_1},
\end{equation}
having angular momentum
\begin{equation}
\mathbf{L}' = \frac{1}{3} \left( \mathbf{D}'_0 + 2 \mathbf{D}'_2 \right)
\end{equation}
and spin
\begin{equation}
\sigma = \boldsymbol{\delta}' \cdot \mathbf{L}',
\end{equation}
and being acted on by a gravitational force that points opposite to
\begin{equation}
\mathbf{G}' = \frac{2}{5} \mathbf{D}'_3. \label{gravity}
\end{equation}
Dropping terms proportional to $\omega$, one then finds that $\boldsymbol{\delta}'$ satisfies the pendulum-like equation
\begin{equation}
\frac{\boldsymbol{\delta}' \times \ddot{\boldsymbol{\delta}}'}{\mu} + \sigma \dot{\boldsymbol{\delta}}' = \mu D_1 \mathbf{G}' \times \boldsymbol{\delta}', \label{fastpendulum}
\end{equation}
warranting the interpretations just given. Furthermore, the length $D_1$ of the pendulum is constant and its motion is restricted by conservation of the energy-like quantity
\begin{equation}
E_D = \mu \mathbf{G}' \cdot \mathbf{D}'_1 + \frac{\mu}{2} \mathbf{L}'^2,
\end{equation}
as well as by a series of more complicated conservation laws. $\mathbf{G}'$ plays the role of gravity but is itself coupled to $\boldsymbol{\delta}'$ directly and to higher moments via $\mathbf{D}'_4$. These features distinguish the fast pendulum from the slow (bipolar) pendulum of Ref.~\cite{hannestad2006}, for which gravity is a fixed external field. Despite the differences, the pendulum analysis accounts for features of fast oscillations seen in linear stability and nonlinear numerics. 

We reiterate, though, that the fast-pendulum analysis relies on vacuum effects being negligible aside from seeding instability. Nonzero $\omega$ leads to nonconservation of $E_D$, $\mathbf{D}_0$, and $D_1$, among other quantities. A different quantity,
\begin{equation}
E_S = \omega \mathbf{B} \cdot \mathbf{S}_0 + \frac{\mu}{2} \left( \mathbf{D}_0^2 - \mathbf{D}_1^2 \right), \label{esenergy}
\end{equation}
is still conserved, however \cite{raffelt2007b}. In isotropic settings, this is the energy of the bipolar pendulum
\begin{equation}
\mathbf{Q} = \mathbf{S}_0 - \frac{\omega}{\mu} \mathbf{B}.
\end{equation}
Letting $\mathbf{q} = \mathbf{Q} / Q$ and assuming isotropic angular distributions, the bipolar pendulum equation reads
\begin{equation}
\frac{\mathbf{q} \times \ddot{\mathbf{q}}}{\mu} + \varsigma \dot{\mathbf{q}} = \omega Q \mathbf{B} \times \mathbf{q}, \label{slowpendulum}
\end{equation}
where the spin $\varsigma$ of the pendulum is
\begin{equation}
\varsigma = \mathbf{q} \cdot \mathbf{D}_0.
\end{equation}
Allowing again for anisotropic angular distributions, the monopole now evolves as
\begin{equation}
\dot{\mathbf{S}}_0 = \omega \mathbf{B} \times \mathbf{D}_0 + \mu \mathbf{D}_0 \times \mathbf{S}_0 - \mu \mathbf{D}_1 \times \mathbf{S}_1, \label{s0eqtn}
\end{equation}
hence
\begin{equation}
\dot{\mathbf{Q}} = \mu \mathbf{D}_0 \times \mathbf{Q} - \mu \mathbf{D}_1 \times \mathbf{S}_1. \label{qeqtn}
\end{equation}
Anisotropy disrupts the slow pendulum, and it does so in a way that is mediated by the fast pendulum $\mathbf{D}_1$.

\begin{figure*}
\centering
\begin{subfigure}{
\centering
\includegraphics[width=.43\textwidth]{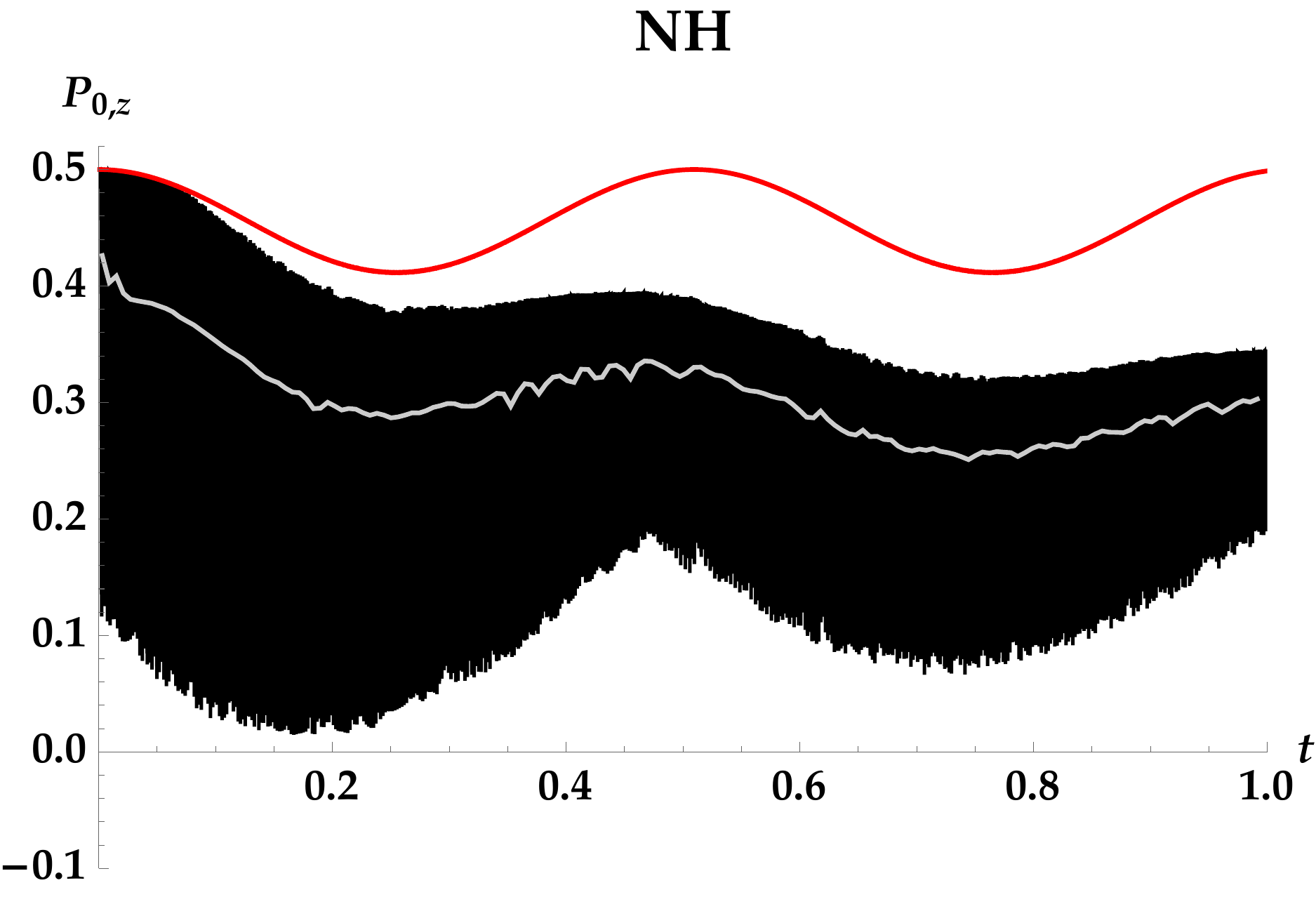}
}
\end{subfigure}
\begin{subfigure}{
\centering
\includegraphics[width=.43\textwidth]{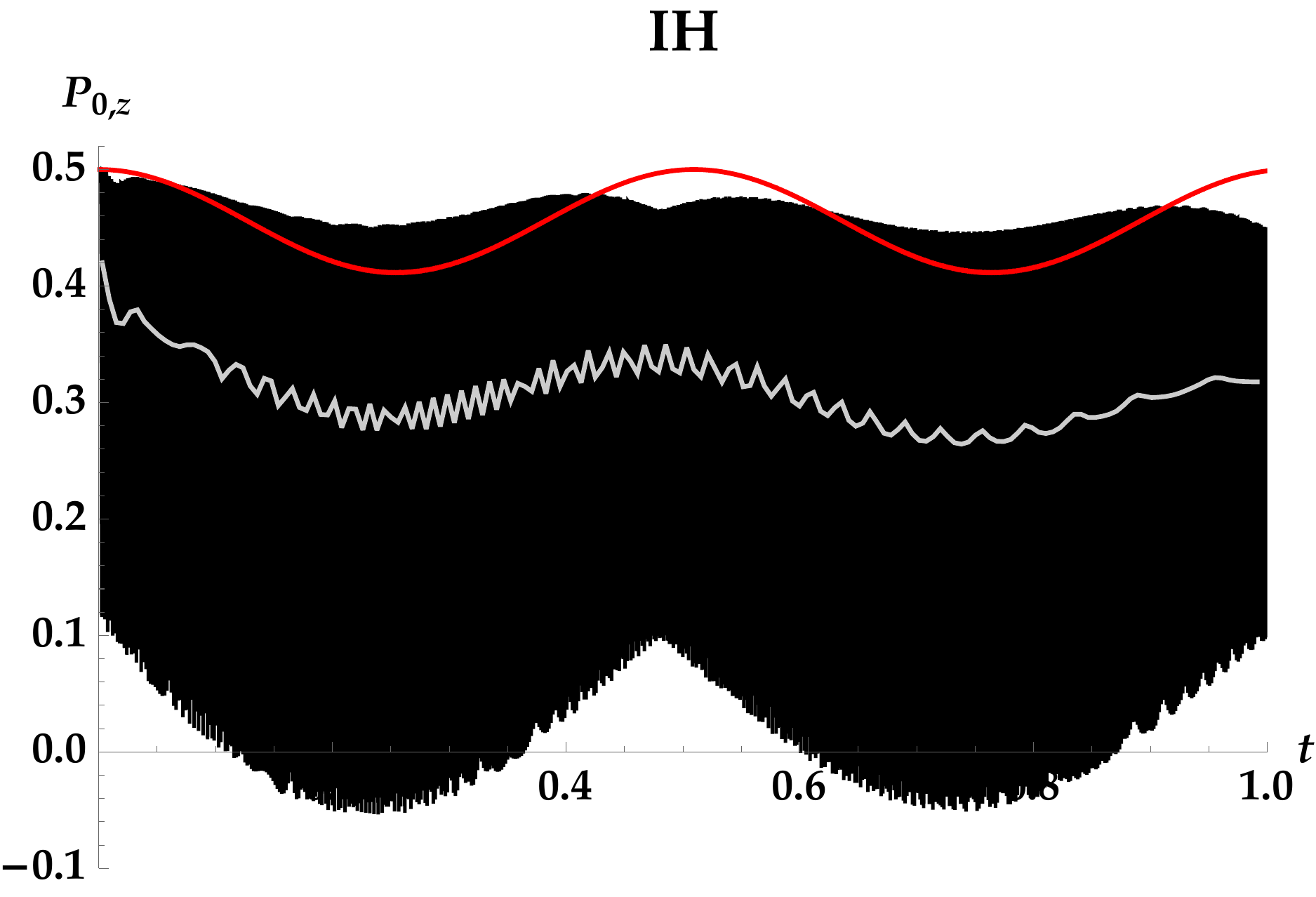}
}
\end{subfigure}

\begin{subfigure}{
\centering
\includegraphics[width=.43\textwidth]{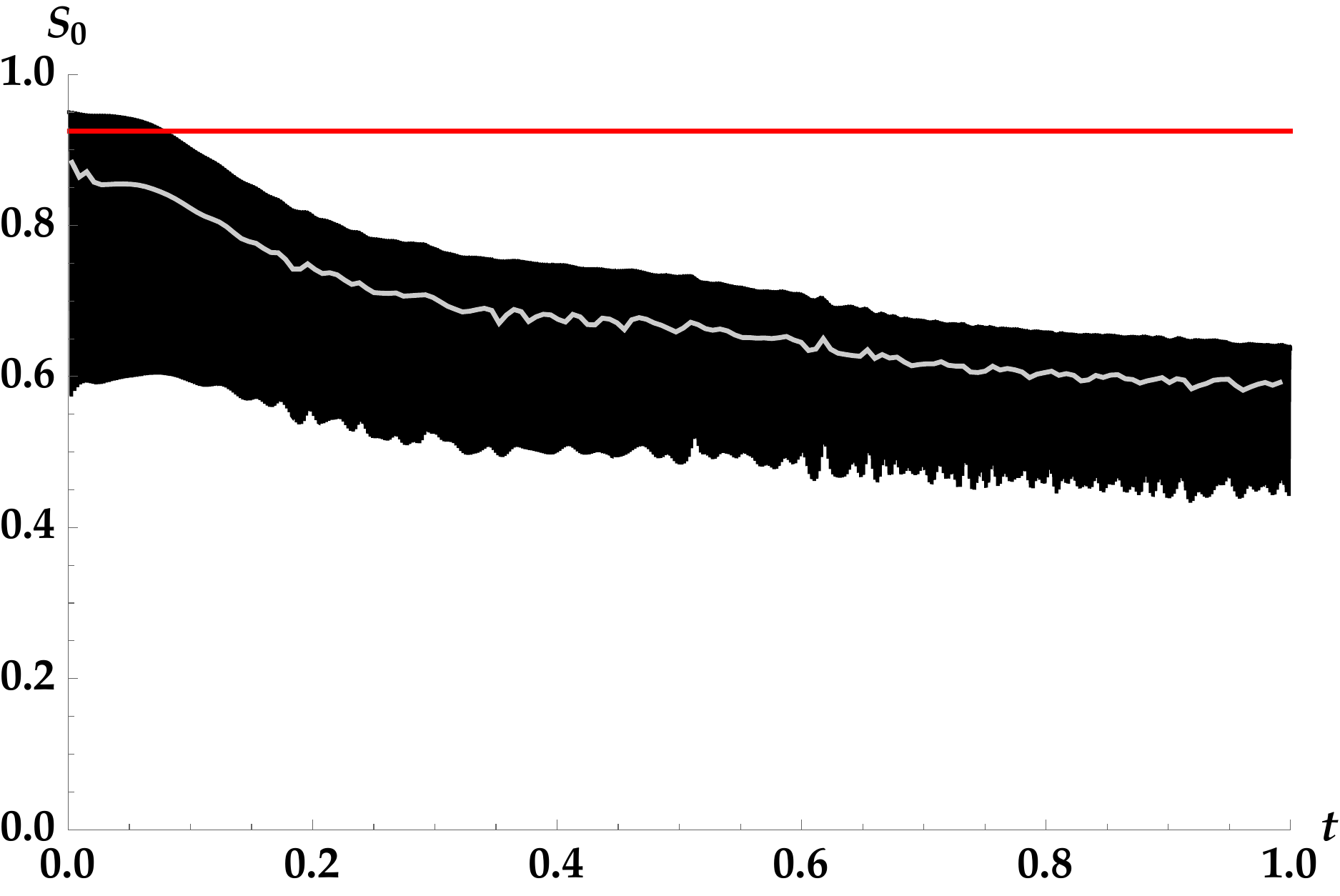}
}
\end{subfigure}
\begin{subfigure}{
\centering
\includegraphics[width=.43\textwidth]{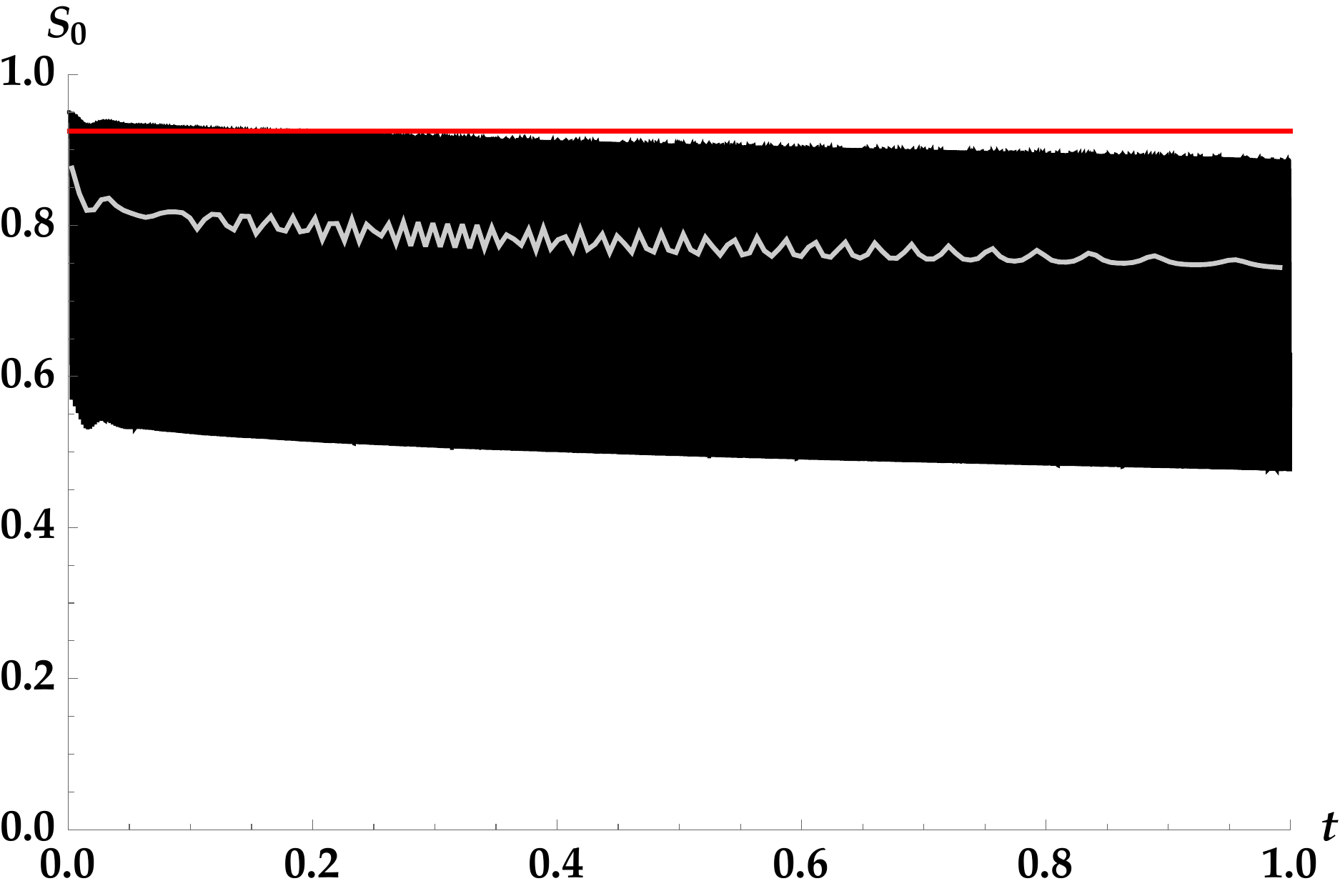}
}
\end{subfigure}

\begin{subfigure}{
\centering
\includegraphics[width=.43\textwidth]{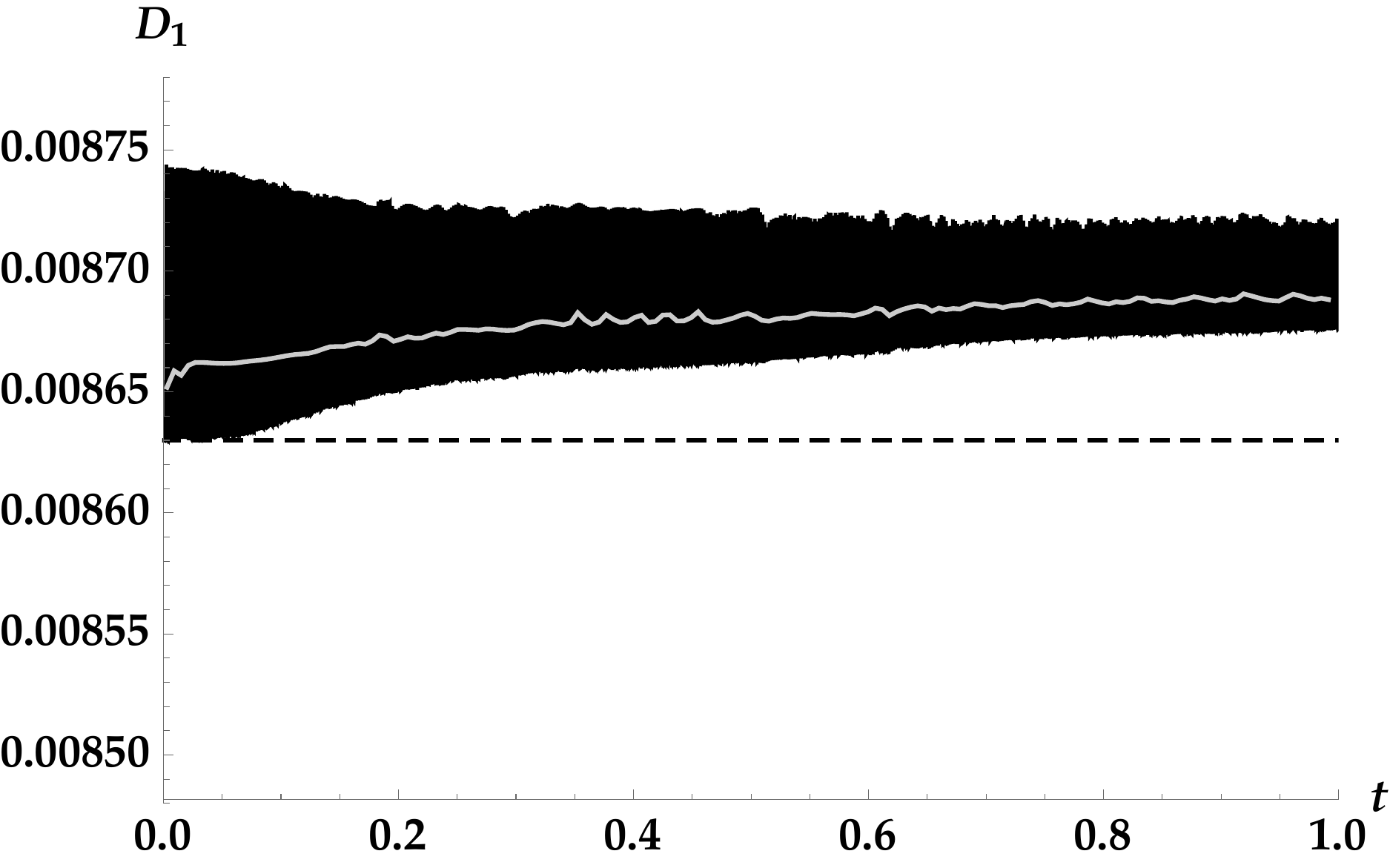}
}
\end{subfigure}
\begin{subfigure}{
\centering
\includegraphics[width=.42\textwidth]{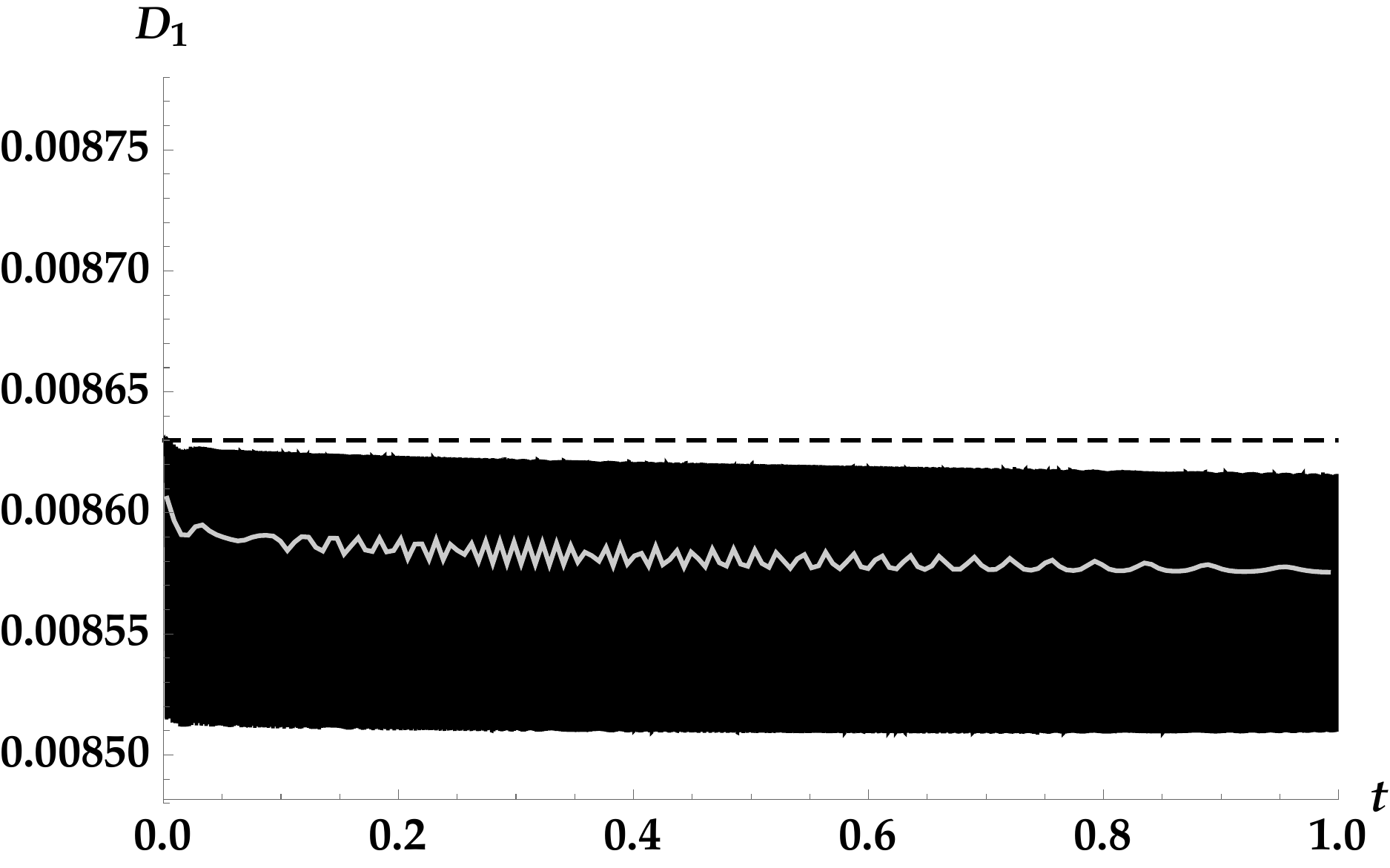}
}
\end{subfigure}
\caption{Flavor evolution---$P_{0,z}$ (top), $S_0$ (middle), and $D_1$ (bottom)---with time $t$ in homogeneous, axially symmetric calculations, using the angular distributions and ratio $\mu / \omega$ from a spherically symmetric \textsc{Fornax} simulation at a radius of 70 km and a post-bounce time of 200 ms. The $\bar{\nu}_e$ angular distribution is rescaled to $\alpha = n_{\bar{\nu}_e} / n_{\nu_e} = 0.90$ (black) and $\alpha = 0.85$ (red), and the normalization is such that $| \mathbf{P}_{0} (0) | = 0.5$. The normal hierarchy (NH) is shown on the left, inverted hierarchy (IH) on the right. Time is in units of $\omega^{-1}$. For typical supernova conditions and a neutrino energy of $\sim 10$ MeV, the horizontal axis spans about a microsecond. Gray curves represent the same data as the black curves but time-averaged over windows of duration $6\times 10^{-3} \omega^{-1}$, spanning about 3 periods of fast collective oscillations. In the bottom panels, the dashed lines mark the initial values. At $\alpha = 0.90$, the system is unstable to fast oscillations, which facilitate kinematic decoherence (decay of $S_0$) by a dephasing mechanism. See text for further discussion.}
\label{p0z_6000m}
\end{figure*}

Raffelt and Sigl \cite{raffelt2007b} studied kinematic decoherence in systems with equal fluxes of $\nu_e$ and $\bar{\nu}_e$. They observed that, because $E_S$ is conserved, $S_0$ can only go to zero if appropriate changes occur in $D_0$, $D_1$, or both. In particular, they identified an exponential solution for $\mathbf{D}_1$ as being key to decoherence. Specifically, when the $\mathbf{D}_l$ vectors are small,
\begin{equation}
\ddot{\mathbf{D}}_1 \cong - \frac{1}{3} \omega \mu \left( \mathbf{B} \cdot \mathbf{S}_0 \right) \mathbf{D}_1. \label{d1exp}
\end{equation}
In the normal hierarchy (NH), $\mathbf{\hat{B}} \cdot \mathbf{\hat{S}}_0 \approx -1$ and exponential growth is possible when $\mathbf{S}_0$ is in its stable position. In the inverted hierarchy (IH), exponential growth only occurs when the pendulum is unstable and swings away from its initial inverted position.

Scenarios with small $\mathbf{D}_l$ vectors are special cases, however, and it has been observed in numerical simulations that collective oscillations often exhibit behavior more in line with single-angle expectations than multi-angle decoherence \cite{duan2006, duan2006b}. As shown in Ref.~\cite{esteban2007}, a large difference between the $\nu_e$ and $\bar{\nu}_e$ fluxes suppresses decoherence. Working with the bulb model, wherein neutrinos are emitted semi-isotropically from a single decoupling surface, and using $F(\nu_\alpha)$ to denote the number flux of flavor $\alpha$, the authors identified
\begin{equation}
\epsilon = \frac{F(\nu_e) - F(\bar{\nu}_e)}{F(\bar{\nu}_e) - F(\bar{\nu}_x)}
\end{equation}
as the decisive asymmetry parameter determining the transition from quasi-single-angle to decoherent multi-angle behavior. Insofar as bulb-model results are accounted for by analyzing the homogeneous Eqs.~\eqref{homogeom}, the suppression of decoherence can be attributed to other terms showing up on the right-hand side of Eq.~\eqref{d1exp}.

It is now understood that the bulb model is itself a special case, assuming as it does that all flavors have angular distributions of the same shape. Quasi-single-angle evolution, kinematic decoherence, and fast oscillations are all related to asymmetries in the fluxes, and in a realistic supernova no single parameter controls the relative significance of each of these facets of the problem. The relevance to fast oscillations of the ratios $D_l / D_0$, for example, was emphasized in Ref.~\cite{johns2020}.

The importance of asymmetry in controlling kinematic decoherence can be seen in the equations
\begin{align}
&\dot{\mathbf{S}}_1 = \omega \mathbf{B} \times \mathbf{D}_1 + \mu \mathbf{D}_0 \times \mathbf{S}_1 + \frac{1}{3} \mu \left( \mathbf{S}_0 + 2 \mathbf{S}_2 \right) \times \mathbf{D}_1, \notag \\
&\dot{\mathbf{D}}_1 = \omega \mathbf{B} \times \mathbf{S}_1 + \frac{2}{3} \mu \left( 2 \mathbf{D}_0 + \mathbf{D}_2 \right)\times \mathbf{D}_1, \label{l1eoms}
\end{align}
where as usual the role of $\mathbf{D}_1$ in coupling adjacent moments is on display. If $D_0$ is large enough, these equations become
\begin{align}
&\dot{\mathbf{S}}_1 \cong \mu \mathbf{D}_0 \times \mathbf{S}_1, \notag \\
&\dot{\mathbf{D}}_1 \cong \frac{4}{3} \mu \mathbf{D}_0 \times \mathbf{D}_1. \label{adiabaticd0}
\end{align}
That is, both $\mathbf{S}_1$ and $\mathbf{D}_1$ tend to track $\mathbf{D}_0$ as it evolves, keeping the decohering term $\mathbf{D}_1 \times \mathbf{S}_1$ small in Eq.~\eqref{s0eqtn}. Then, since
\begin{equation}
\dot{\mathbf{D}}_0 = \omega \mathbf{B} \times \mathbf{S}_0
\end{equation}
as always, the equation of motion of $\mathbf{S}_0$ is approximately that of the isotropic bipolar system. The larger $D_0$ is, the more adiabatic the tracking, the more suppressed the decoherence, and the more accurate the isotropic approximation. A large isotropic asymmetry $D_0$ is not the only way to prevent multi-angle decoherence, however. Large initial values of $D_1$ can also preempt the exponential solution of Eq.~\eqref{d1exp}.

The goal of the analysis thus far has been to indicate how fast collective oscillations, slow collective oscillations, and multi-angle decoherence all arise out of the same system of equations. These phenomena can be understood in isolation by appealing to Eqs.~\eqref{fastpendulum}, \eqref{slowpendulum}, and \eqref{d1exp}, respectively. In a linear analysis, they correspond to the fast, bipolar, and multi-zenith-angle instabilities.

\subsection{Relaxation and cascade}

We now turn to one of our central points, which is that decoherence need not proceed through exponential decay of $S_0$. Even in the quasi-isotropic limit, $\mathbf{D}_1 \times \mathbf{S}_1$ does not vanish precisely and decoherence is expected to take place at some level. Moreover, fast oscillations of $\mathbf{D}_1$ have the potential to accelerate the relaxation rate by dephasing $\mathbf{D}_1$ and $\mathbf{S}_1$. This is quite in contrast to bipolar oscillations, which tend to keep these vectors (anti)aligned as discussed above.

Acceleration of kinematic decoherence by FFC is shown in Fig.~\ref{p0z_6000m}, which presents numerical solutions using $\mu / \omega$ and angular distributions taken from a spherically symmetric \textsc{Fornax} simulation at 200 ms post-bounce and a radius of $70$ km. The number density $n_{\bar{\nu}_e}$ is treated as a free parameter, and the ratio $\alpha = n_{\bar{\nu}_e} / n_{\nu_e}$ is varied in order to compare evolution under conditions stable to FFC ($\alpha = 0.85$) and conditions unstable to it ($\alpha = 0.90$).

The top panel of the figure compares the isotropic flavor composition $P_{0,z}$ for the two choices of $\alpha$. In both mass hierarchies, $\alpha = 0.85$ exhibits quasi-isotropic evolution, whereas $\alpha = 0.90$ exhibits fast oscillations modulated by bipolar motion and---as confirmed by the middle panel, showing $S_0$---by kinematic decoherence. Consistent with the foregoing analysis, no decoherence is visible in the quasi-isotropic evolution, but it is substantial in the NH when the system is unstable to fast oscillations. FFC-assisted decoherence is evident in the IH as well, albeit to a lesser extent.

The bottom panel indicates that decoherence proceeds through the growth (NH) or decay (IH) of $D_1$, as expected from the conservation of $E_S$. (Though not shown, $D_0$ is very nearly constant in all calculations presented in the figure.) On $\mu^{-1}$ timescales and in the limit $\mu \gg \omega$, the mass hierarchies are described by approximately identical equations of motion. But as noted previously, $D_1$ is constant under the same assumptions. Decoherence and its hierarchy-dependence can thus be traced back to the term $\omega \mathbf{B} \times \mathbf{S}_1$ in the second of Eqs.~\eqref{l1eoms}. In particular,
\begin{equation}
\frac{d}{dt} D_1^2 = - 2 \omega \mathbf{B} \cdot \left( \mathbf{D}_1 \times \mathbf{S}_1 \right).
\end{equation}
The mass hierarchies are distinguished by the orientation of the decohering term $\mathbf{D}_1 \times \mathbf{S}_1$ with respect to $\omega \mathbf{B}$.

\begin{figure*}
\centering
\begin{subfigure}{
\centering
\includegraphics[width=.45\textwidth]{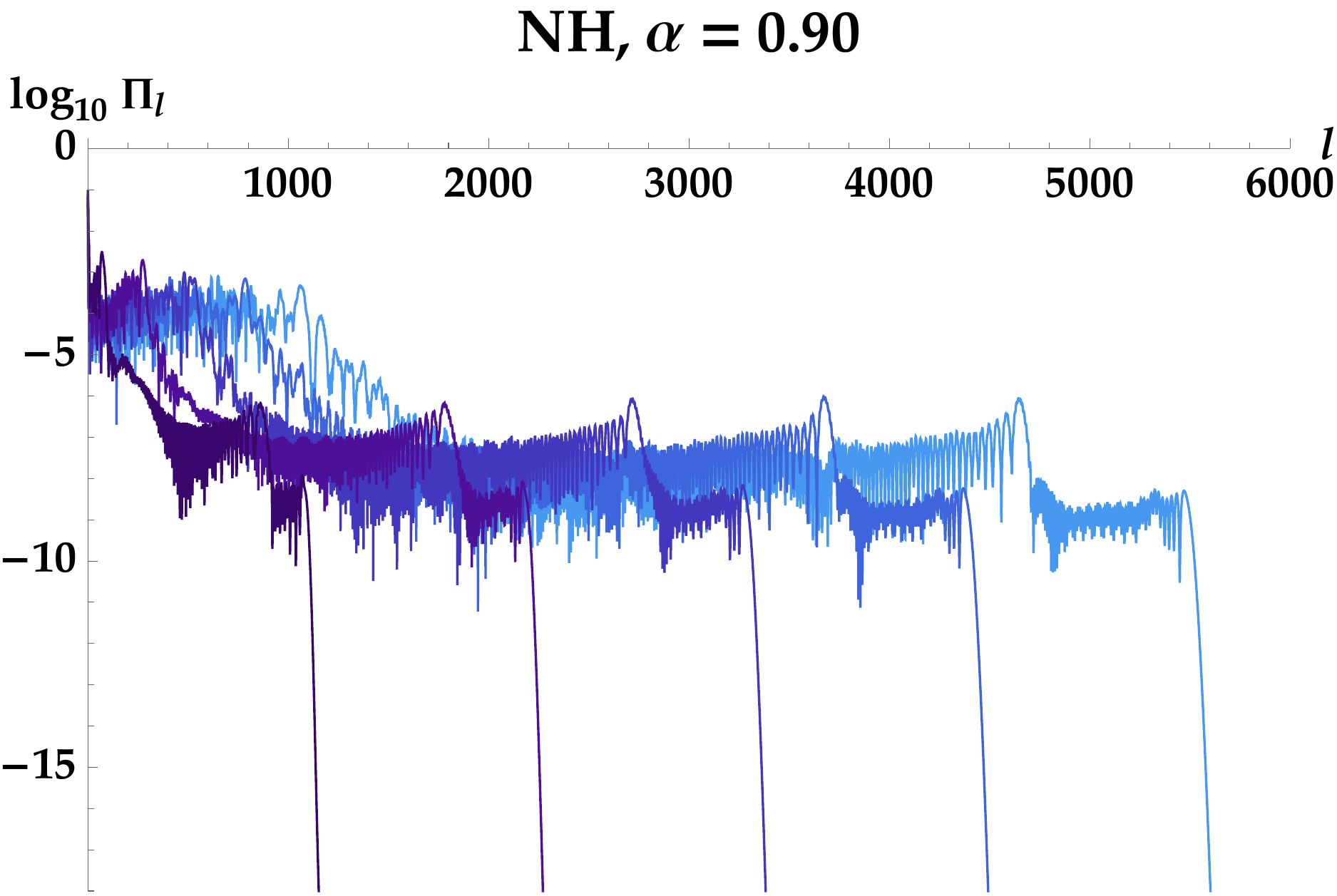}
}
\end{subfigure}
\begin{subfigure}{
\centering
\includegraphics[width=.45\textwidth]{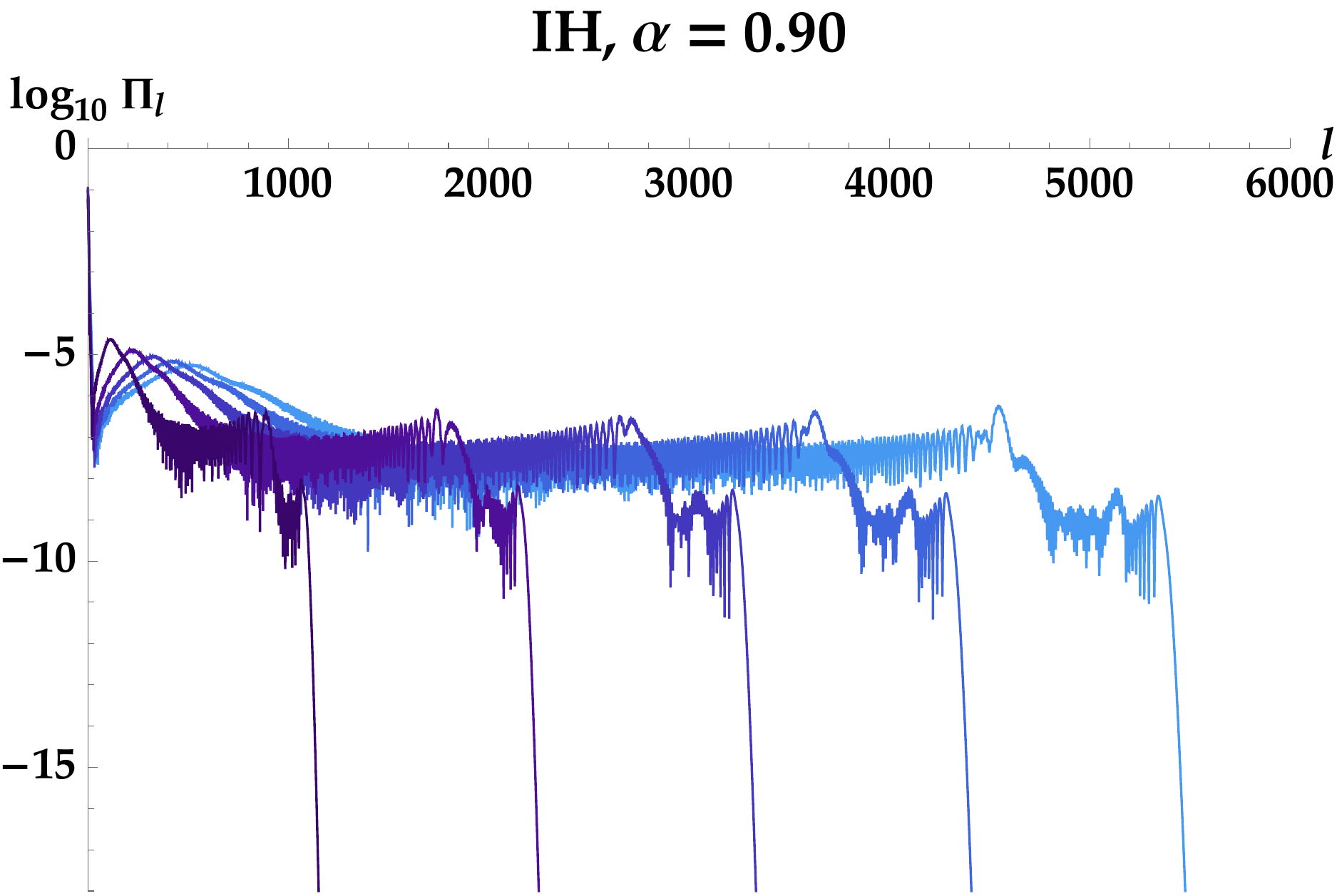}
}
\end{subfigure}

\begin{subfigure}{
\centering
\includegraphics[width=.45\textwidth]{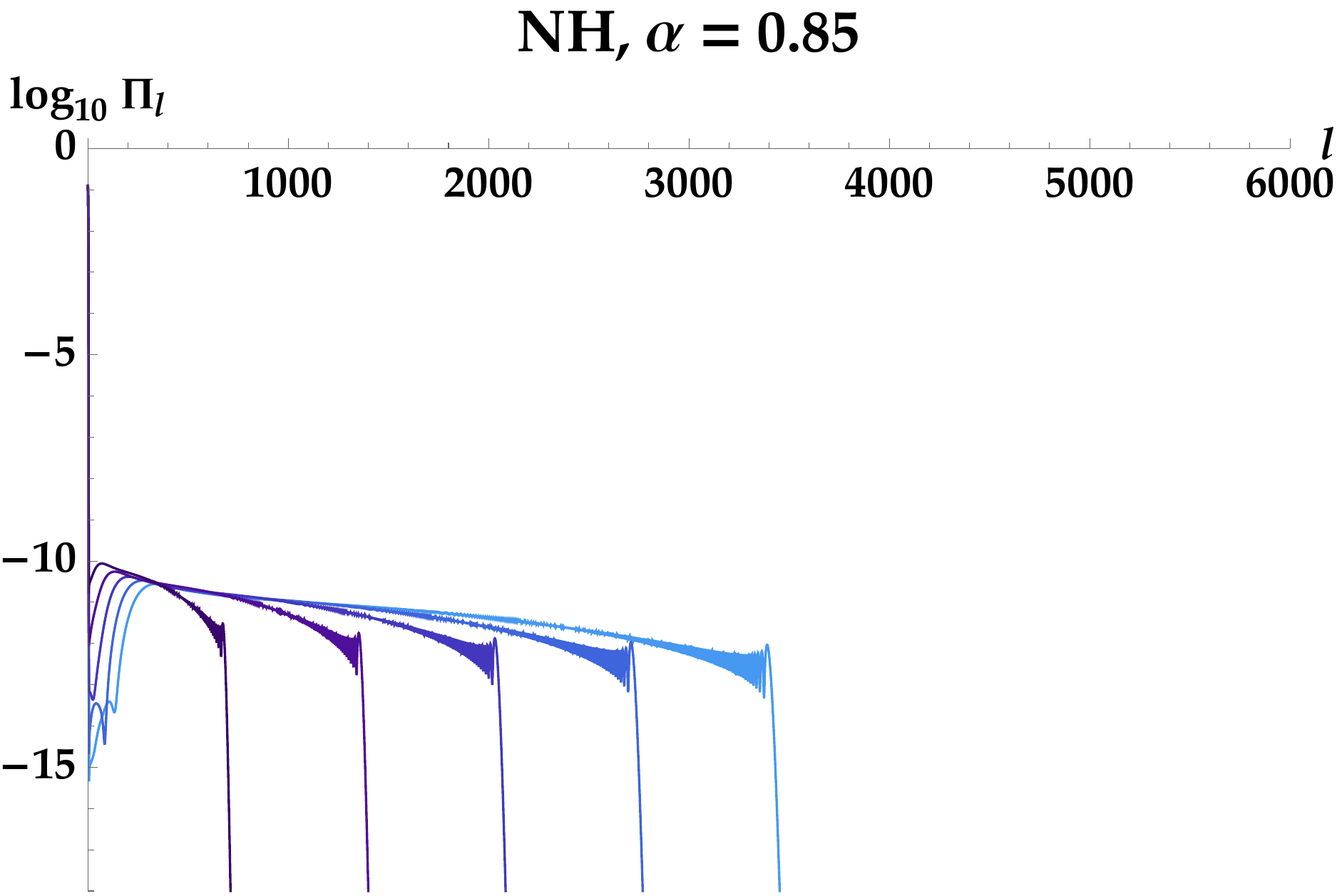}
}
\end{subfigure}
\begin{subfigure}{
\centering
\includegraphics[width=.45\textwidth]{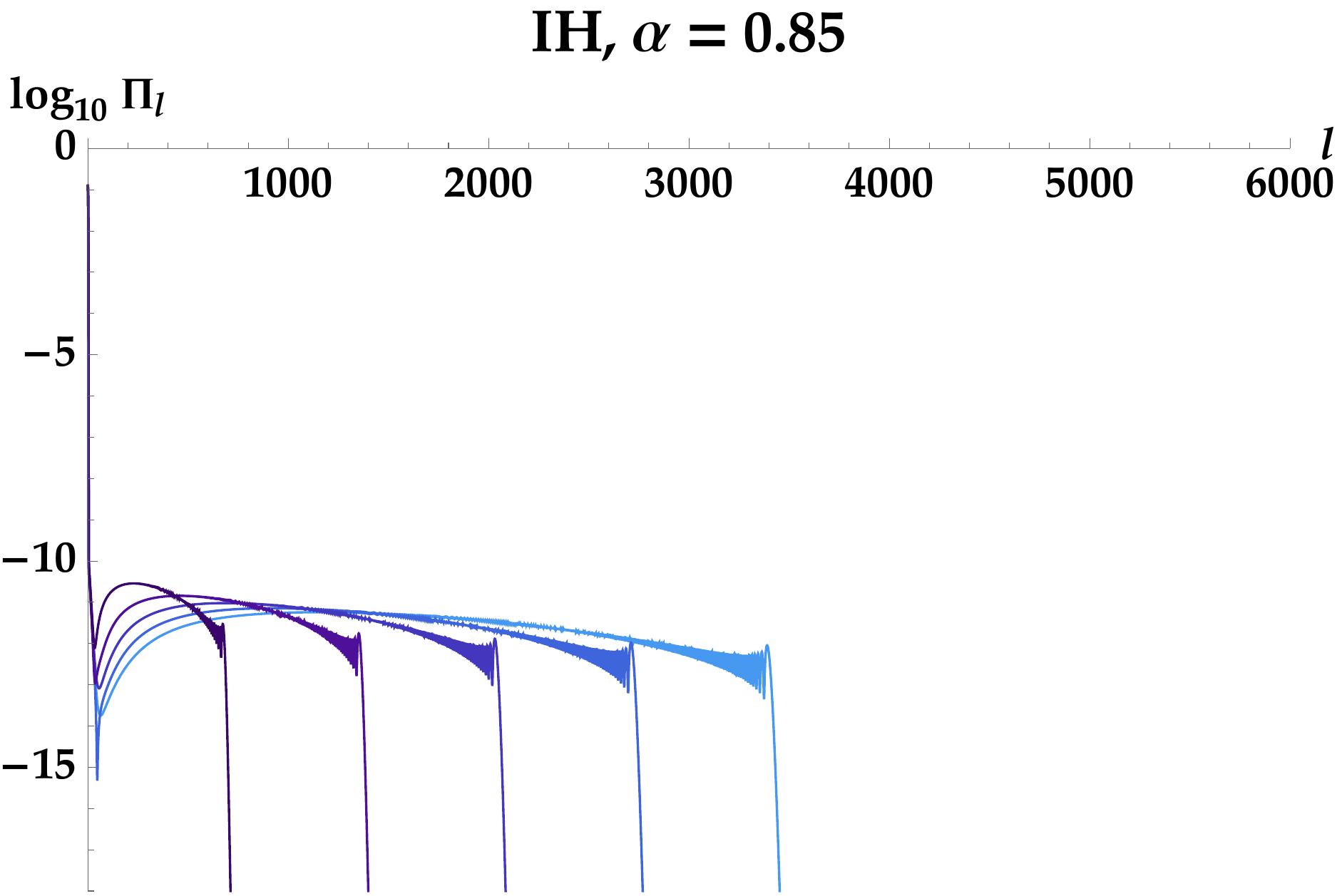}
}
\end{subfigure}
\caption{Momentum-space angular power spectra of neutrino flavor for the same set of calculations presented in Fig.~\ref{p0z_6000m}. The power $\Pi_l$ in angular moment $l$ is defined in Eq.~\eqref{powerdef}. From darkest to lightest, the curves show $\log_{10} \Pi_l$ at $t = 0.2$, $0.4$, $0.6$, $0.8$, and $1.0$, in units of $\omega^{-1}$. FFC enhances cascade and hastens relaxation. All cases show interesting features at angular scales intermediate between the expanding front of nonzero power and the low-$l$ scales directly involved in fast and bipolar collective oscillations.}
\label{pilvsl_6000m}
\end{figure*}

\begin{figure*}
\centering
\begin{subfigure}{
\centering
\includegraphics[width=.31\textwidth]{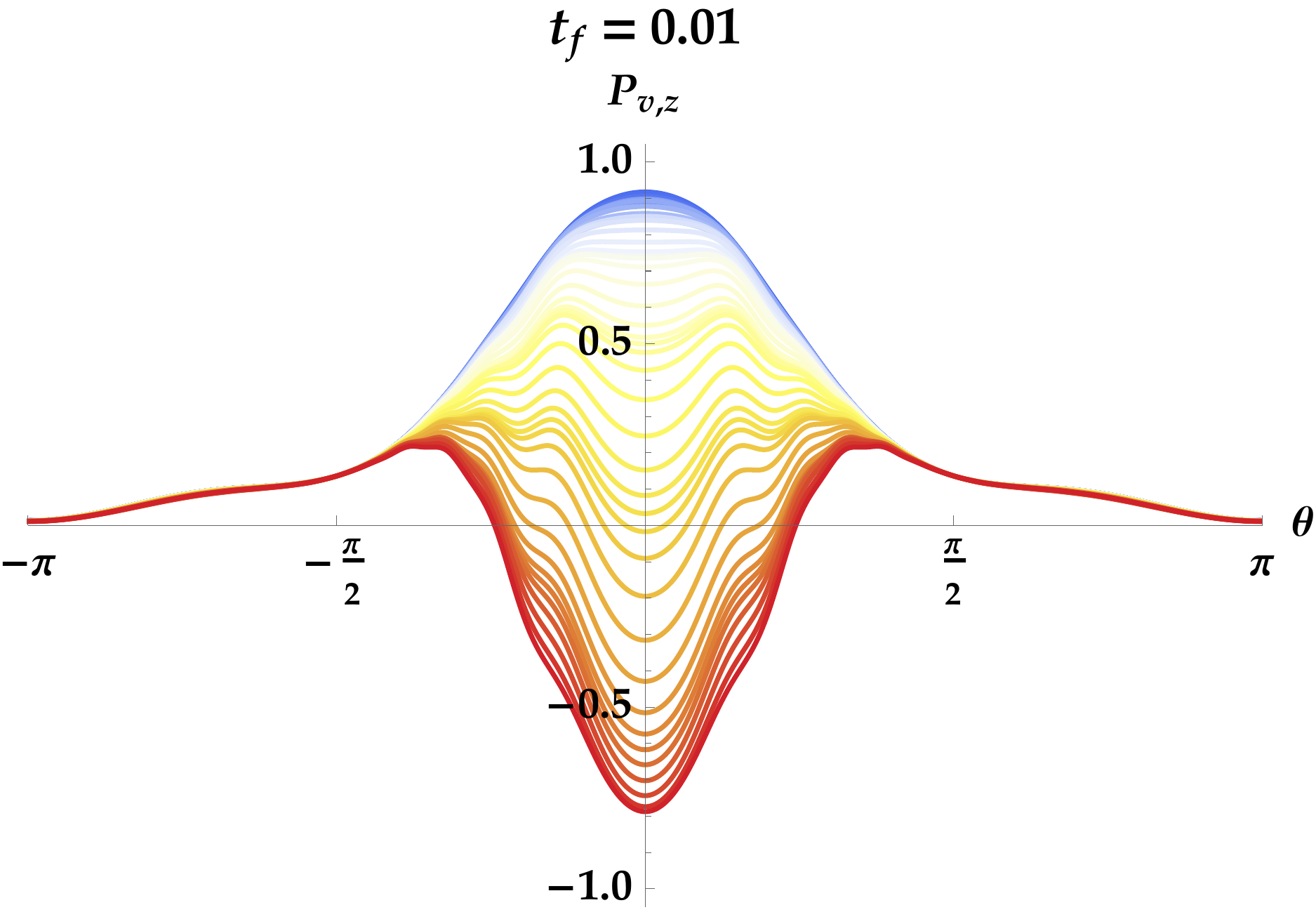}
}
\end{subfigure}
\begin{subfigure}{
\centering
\includegraphics[width=.31\textwidth]{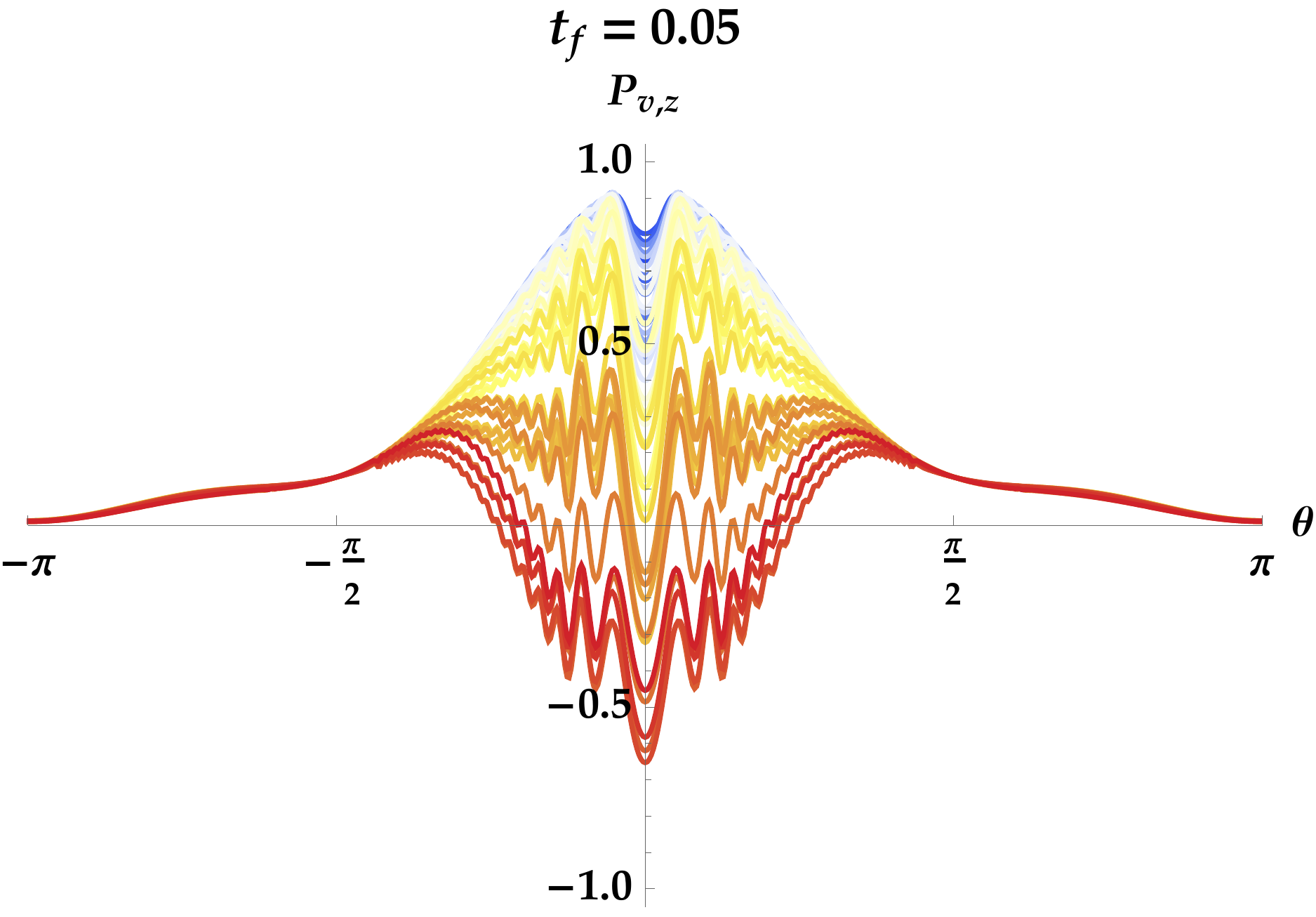}
}
\end{subfigure}
\begin{subfigure}{
\centering
\includegraphics[width=.31\textwidth]{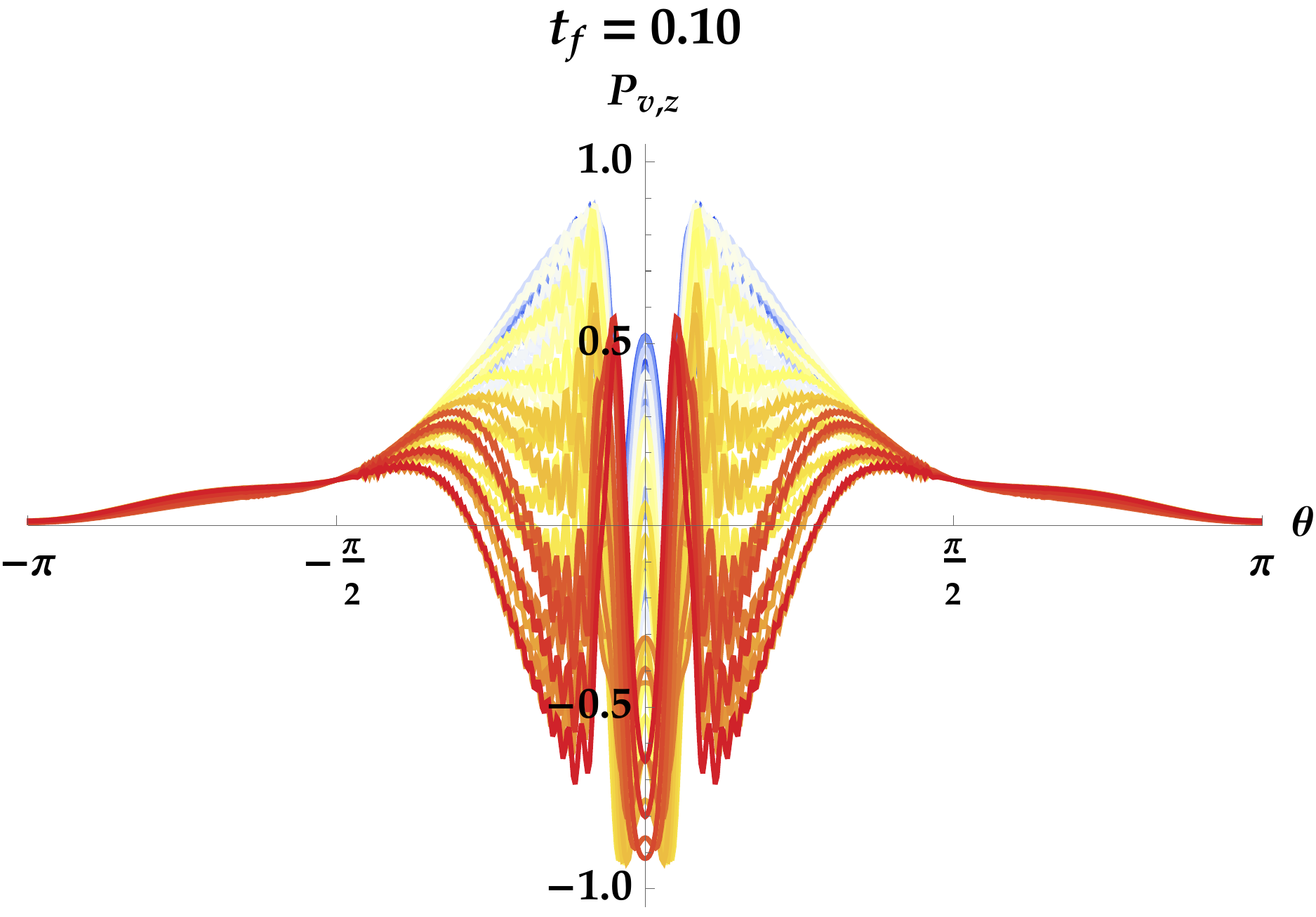}
}
\end{subfigure}
\caption{Flavor composition $P_{v,z}$ plotted as a function of propagation angle $v = \cos \theta$, for the $\alpha = 0.90$, NH calculation of the previous figures. The curves are color-coded by time, beginning at the top of a fast-oscillation dip (blue), lasting a duration of  $\sim 10^{-3}$, and ending at the bottom of the dip at time $t_f$ (red). As usual, times are given in units of $\omega^{-1}$. Finite-$\omega$ effects---the disruption of the fast pendulum and the cascade of power to smaller angular scales---become increasingly apparent at later times.}
\label{pvz_6000m}
\end{figure*}

Conservation of the energy-like quantity $E_S$ offers one perspective on collisionless relaxation. Another perspective is provided by unitarity. Defining the power $\Pi_l$ in angular moment $l$,
\begin{equation}
\Pi_l = \left( l + \frac{1}{2} \right) \big| \mathbf{P}_l \big| ^2, \label{powerdef}
\end{equation}
unitarity implies that the sum of the power over all angular scales is conserved:
\begin{equation}
\sum_{l=0}^\infty \Pi_l = \textrm{const}. \label{unitarity}
\end{equation}
(Similar quantities can of course be defined for antineutrinos. In the cases we study the evolution of antineutrinos is comparable to that of neutrinos.) As the isotropic moment relaxes, higher ones are excited. Recalling that the term in the Hamiltonian proportional to $\mathbf{D}_1$ couples each $\mathbf{P}_l$ to its nearest neighbors, the default expectation is that power will continue to be relayed out to higher $l$. This process constitutes collisionless relaxation via phase-space transfer in homogeneous, axially symmetric flavor evolution.

Power is principally lost from the isotropic moments. Although we do not analyze the impact of collisions in this paper, their influence---and especially the relative importance of emission/absorption and scattering processes---will depend on where power resides and how it is transferred collisionlessly.

Fig.~\ref{pilvsl_6000m} displays the ``angular power spectra'' in both mass hierarchies, again comparing $\alpha = 0.85$ and $\alpha = 0.90$. As anticipated, cascade occurs in all cases and is greatly magnified by FFC. Enhancement is seen both in the rate at which power travels out to higher $l$ and in the amplitude at which it does so. As the cascade front moves outward, roughly flat regions form in its wake, expanding and becoming flatter with time. (For $\alpha = 0.90$, the rapid oscillations as a function of $l$ are flattened by time-averaging over many fast oscillation periods.) The orders-of-magnitude difference between typical $\Pi_l$ values in the $\alpha = 0.90$ and $\alpha = 0.85$ calculations explains why kinematic decoherence is visible in Fig.~\ref{p0z_6000m} for the former but not the latter.

These regions are analogous to the inertial ranges encountered in fluid turbulence, which span intermediate scales between those at which driving and dissipation occur. Here the driving is oscillations at large angular scales. Oscillations induce dephasing and thereby momentum-space transfer. In this collisionless system, there is no analogue of a dissipation scale, and so power continues to cascade perpetually out to higher $l$. The asymptotic state, one imagines, is infinitesimal power equally distributed over all moments out to infinity. This is the closest the system can get to fully relaxed while still satisfying unitarity, Eq.~\eqref{unitarity}. 

Over time, as power cascades to higher $l$, small-scale angular features becomes increasingly apparent in the flavor composition $P_{v,z}$ as a function of propagation angle $v$. Fig.~\ref{pvz_6000m} shows this development by comparing the $v$-dependence of fast oscillations at different times. Interaction of FFC with momentum-space transfer and slow collective evolution causes the $P_{v,z}$ profile and its periodicity to be increasingly disrupted \cite{johns2020, shalgar2020}. The effect should not be overstated, however, as the outline of the $\omega \rightarrow 0$ collective behavior persists to some degree, at least in these calculations. The essence of the fast pendulum remains even as the pendulum is affected and modulated by finite-$\omega$ effects.

We have seen, in this subsection, how the dephasing mechanism of Sec.~\ref{analysis} brings about kinematic decoherence; how the mechanism is intensified by FFC; how the power lost at large angular scales cascades down to smaller ones, relaxing the system in a unitary manner; and how cascade manifests as small-scale angular structures in the flavor evolution. Next we look more closely at the nature of momentum-space transfer far from the monopole.

\subsection{Transfer at small angular scales}

Further insight into the momentum-space dynamics is gained by observing the evolution of an initially isolated seed in the angular power spectrum. An experiment of this kind allows us to focus in on transfer without the complicating features specific to low-$l$ evolution, namely fast and bipolar collective oscillations and the existence of a natural cutoff at $l=0$. (In a real physical setting these features are of course essential, and Ref.~\cite{johns2020} was devoted to studying them.) The section following this one then goes to the other extreme, examining the effects on transfer of an artificial cutoff at $l = l_\textrm{max}$.

\begin{figure}
\centering
\begin{subfigure}{
\centering
\includegraphics[width=.44\textwidth]{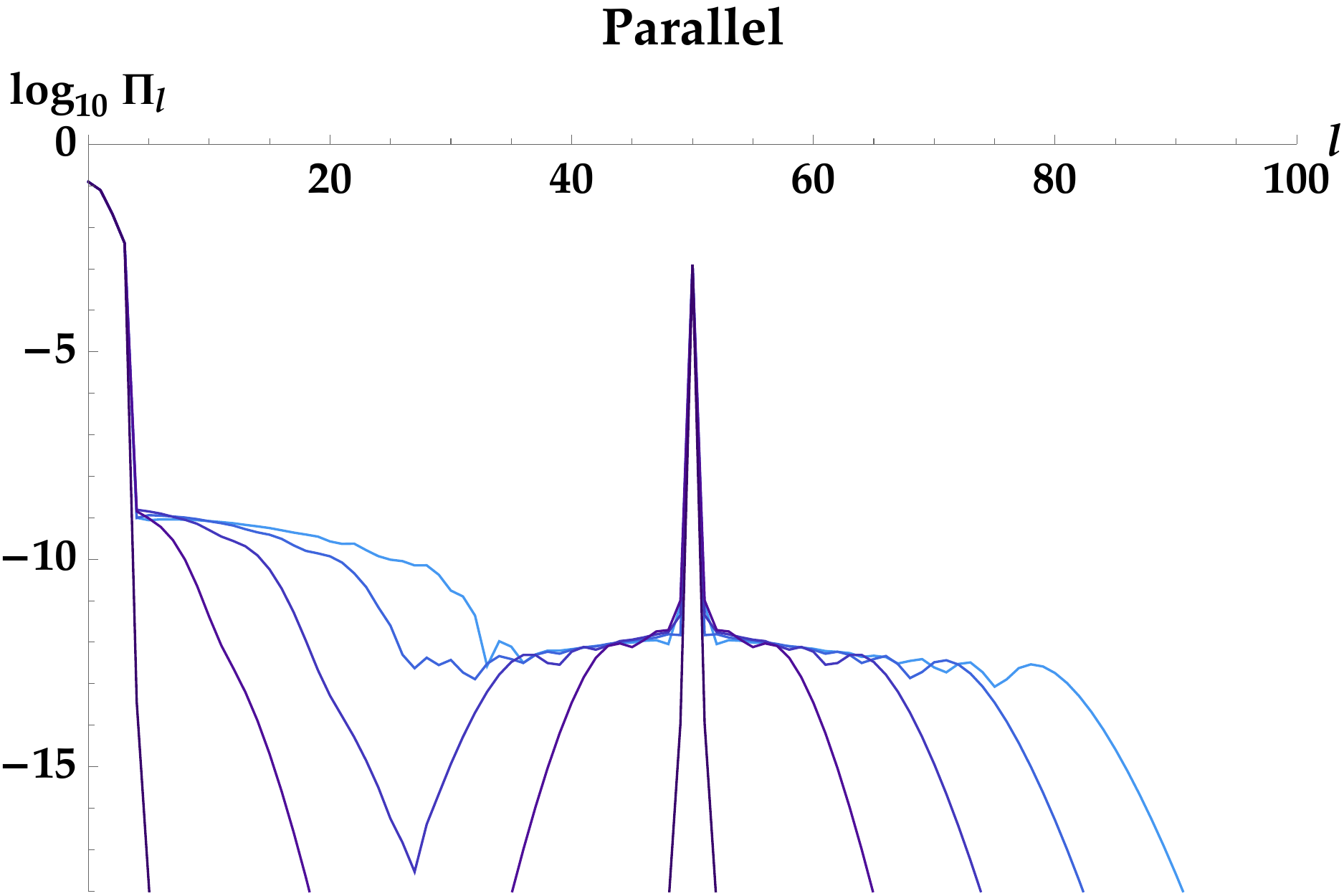}
}
\end{subfigure}
\begin{subfigure}{
\centering
\includegraphics[width=.44\textwidth]{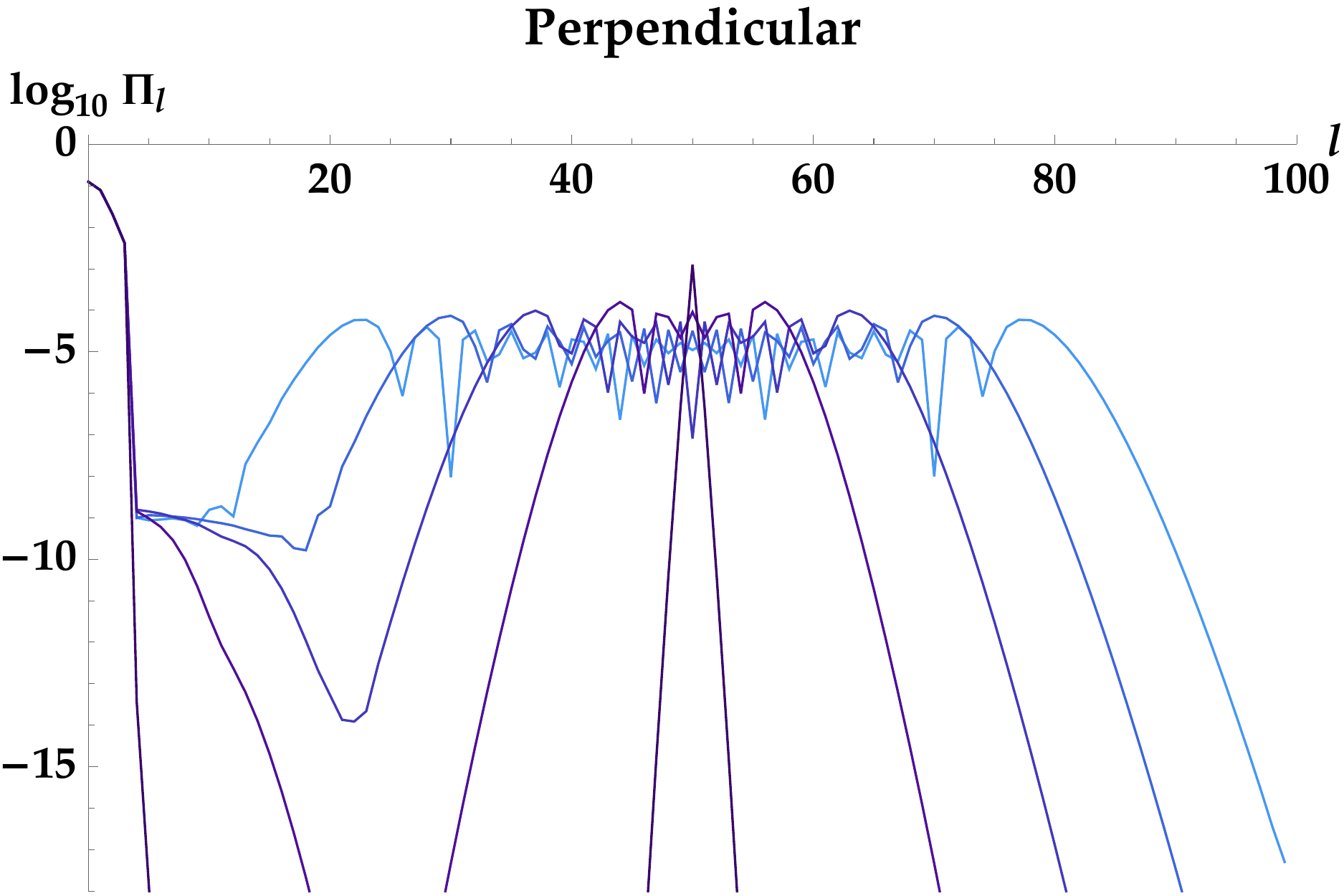}
}
\end{subfigure}
\caption{Angular power spectra in calculations with power seeded at $l=50$. In the plot labels, ``Parallel'' indicates that $\mathbf{P}_{50}$ and $\mathbf{\bar{P}}_{50}$ are set parallel to the flavor axis $\mathbf{z}$ at $t=0$; ``Perpendicular'' indicates that they are perpendicular to the flavor axis and parallel to each other. From darkest to lightest, the curves are at times $1.1 \times 10^{-5}$, $2.2 \times 10^{-3}$, $4.4 \times 10^{-3}$, $6.6 \times 10^{-3}$, and $8.8 \times 10^{-3}$ in units of $\omega^{-1}$.}
\label{pilvsl_diff}
\end{figure}

In Fig.~\ref{pilvsl_diff}, power has been placed by hand at $t=0$ in the $l=50$ moments of $\mathbf{P}_v$ and $\mathbf{\bar{P}}_v$. Color-coded snapshots of the angular power spectrum are shown. These calculations were done using the NH and $\alpha = 0.85$. Without the $l=50$ seeds, they would be the same as the calculation shown in the bottom-left panel of Fig.~\ref{pilvsl_6000m}.

The top panel of Fig.~\ref{pilvsl_diff} shows the evolution that results when the seeds are chosen to be parallel to the flavor axis $\mathbf{z}$ at $t=0$. Gently sloping plateaus form on the sides of the $l=50$ spike in the angular power spectrum and expand outward over time. Because of the relatively low amplitudes of these plateaus, little power is lost from $l=50$.

The bottom panel shows the evolution when the seeds are initially \textit{perpendicular} to the flavor axis. In this case, a single plateau forms, which encompasses $l=50$ and is flat when averaged over suitable intervals of time. The entire spike vanishes into this expanding region, which consequently has a much higher amplitude than in the top panel, where only a tiny fraction of the seeded power is sapped by the moments neighboring $l=50$. In contrast with the top panel, here the inverse cascade emanating from the seed overwhelms the direct cascade from low $l$.

The plateaus in $\Pi_l$ form due to the final term in
\begin{align}
&\dot{\mathbf{P}}_l = \omega \mathbf{B} \times \mathbf{P}_l + \mu \mathbf{D}_0 \times \mathbf{P}_l \notag \\
&\hspace{.85 in}- \frac{\mu}{2} \mathbf{D}_1  \times \left( a_l \mathbf{P}_{l-1} + b_l \mathbf{P}_{l+1} \right). \label{pleqtn}
\end{align}
$\mathbf{D}_1$ causes $\mathbf{P}_{49}$ and $\mathbf{P}_{51}$ to sap power from the seed, but only when $\mathbf{P}_{50}$ has a part orthogonal to $\mathbf{D}_1$. This accounts for the different behaviors seen in the two panels. When $\mathbf{\hat{P}}_{50}$ starts (and stays) close to $\pm \mathbf{\hat{D}}_1$, power gradually leaks out of $l=50$, but when they are initially orthogonal, $\mathbf{D}_1$ quickly smears out the power concentrated there.

Based on a non-exhaustive parametric study with fixed initial distributions, we find that the spreading rate and height of the plateau are proportional to $\mu$ and, in the parallel case, $\left( \omega \sin 2\theta / \mu \right)^2$, respectively. (Since the seed quickly dissolves into the plateau in the perpendicular calculation, the plateau's height is simply set by the seed power and the plateau width.) These findings are sensible. The final term in Eq.~\eqref{pleqtn} tells us that the rate of moment transfer is scaled by $\mu$, hence the proportionality to $\mu$ of the spreading rate. As for the height, the angular separation between $\mathbf{\hat{P}}_{50}$ and $\pm \mathbf{\hat{D}}_1$ is roughly expected to scale like $\omega \sin 2 \theta / \mu$, where the numerator is the typical separation due to vacuum oscillations and the denominator is suppression due to vectors adiabatically tracking $\mathbf{D}_0$, as discussed below Eq.~\eqref{adiabaticd0}. Because $\Pi_l \propto | \mathbf{P}_l |^2$, one roughly expects the scaling reported above.

Unsurprisingly, larger initial $D_0$, which enhances adiabaticity, does tend to decrease the plateau height. Conversely, a larger initial $D_1$ \textit{increases} the height, as it shifts the balance from adiabaticity to decoherence: The third term on the right-hand side of Eq.~\eqref{pleqtn} is enhanced relative to the second one. The empirical scalings are not always straightforward, however, as the typical angular separations are determined by competing effects. The spreading rate, dependent only on the last term of Eq.~\eqref{pleqtn}, is independent of $D_0$ and proportional to $D_1$.

In the parallel case, the decoherence rate of $\mathbf{P}_{50}$ is expected to be roughly the product of the height and twice the rate at which each plateau front advances, the factor of two coming from the fact that the seed sources both a direct and an inverse cascade. As noted before, these plateaus slope downward away from $l=50$. We have not attempted to estimate the slope. In the perpendicular case, the plateau is approximately flat when averaged over a time scale long relative to the fluctuations in a given $\mathbf{P}_l$ and short relative to the expansion rate. The flatness reflects the efficient scrambling of the initial seed. The $\mu \mathbf{D}_1$ term that causes the plateau to expand is, naturally, also responsible for transferring power between neighbors \textit{within} the plateau. As cascade proceeds, power therefore remains equilibrated (again in a time-averaged sense) over the entirety of the plateau. The parallel seed, in contrast, is a persistent, ``low-entropy'' feature that greatly delays this sort of equilibration.

It is worth making contact here with the observation of Ref.~\cite{raffelt2007b} that the equations of motion bear resemblance to drift--diffusion equations in multipole space. Treating $l$ as a continuous variable, the authors showed that the $\mathbf{D}_l$ equation of motion, for example, can be rewritten as [their Eq.~(50)]
\begin{align}
\dot{\mathbf{D}}_l \cong \omega \mathbf{B} &\times \mathbf{S}_l - \mu \mathbf{D}_1 \times \left( \frac{1}{2l+1} \frac{d \mathbf{D}_l}{dl}+ \frac{1}{2} \frac{d^2 \mathbf{D}_l}{dl^2} \right) \notag \\
&\hspace{.3 in}+ \mu \left( \mathbf{D}_0 - \mathbf{D}_1 \right) \times \mathbf{D}_l. \label{drift}
\end{align}
In this interpretation of the equations, equipartition between the kinetic and potential energies constituting $E_S$ [Eq.~\eqref{esenergy}] implies that the drift speed and diffusivity are proportional to $\sqrt{\omega \mu}$. Equipartition was expected and observed in Ref.~\cite{raffelt2007b} due to saturation of the exponential instability. The situation here is different. Because our test cases are stable against the exponentially growing $\mathbf{D}_1$ solution, equipartition is not reached, and relaxation occurs instead through subtle dephasing effects in which the oscillatory behavior of $\mathbf{D}_1$ is critical. The drift--diffusion behavior reflects this distinction, as it is linked to the mechanism that releases power from low $l$.

%Our observations in this subsection are consistent with the finding of the previous one that FFC can enhance both the speed and amplitude of cascade. Fast oscillations of $\mathbf{D}_1$ drive the development of perpendicular components, increasing the plateau height. Moreover, fast-unstable conditions typically require $D_1 / D_0$ to be sufficiently large \cite{johns2020}. As discussed already, increasing this ratio has the effect of shifting the balance away from $\mathbf{D}_0$-dominated adiabatic evolution, and increasing $D_1$ specifically has the effect of amplifying the transfer rate. 

\begin{figure*}
\centering
\begin{subfigure}{
\centering
\includegraphics[width=.43\textwidth]{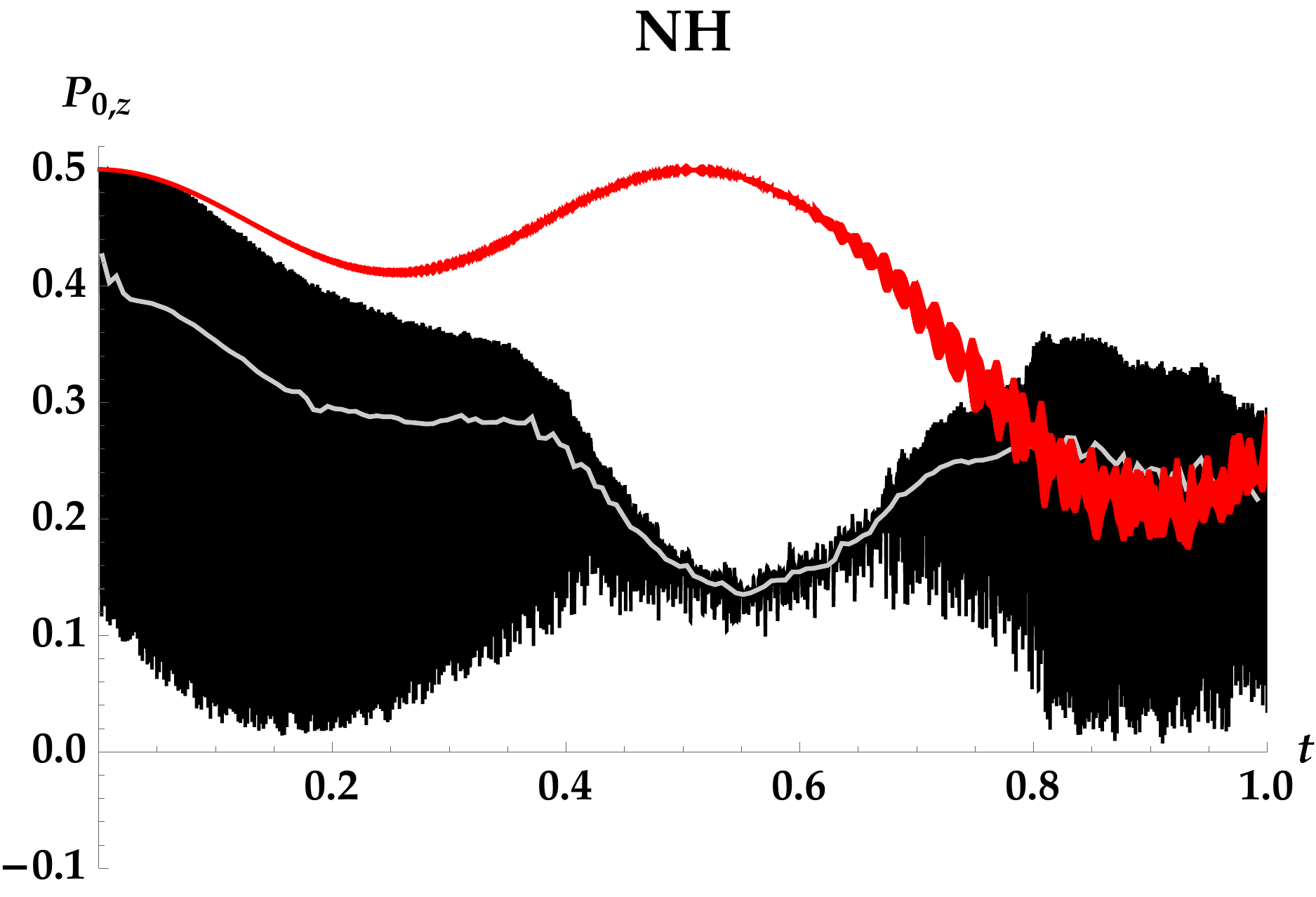}
}
\end{subfigure}
\begin{subfigure}{
\centering
\includegraphics[width=.43\textwidth]{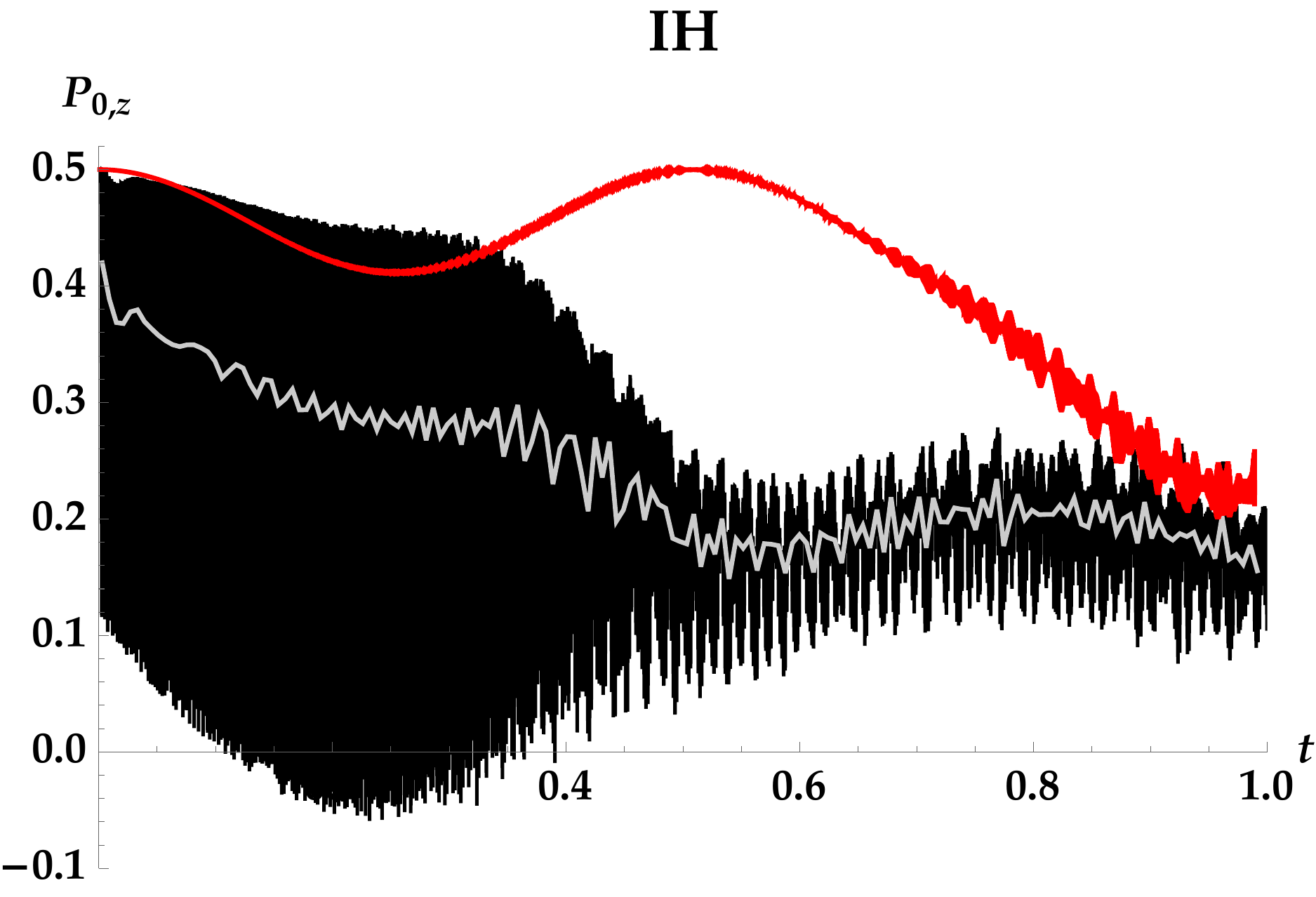}
}
\end{subfigure}

\begin{subfigure}{
\centering
\includegraphics[width=.43\textwidth]{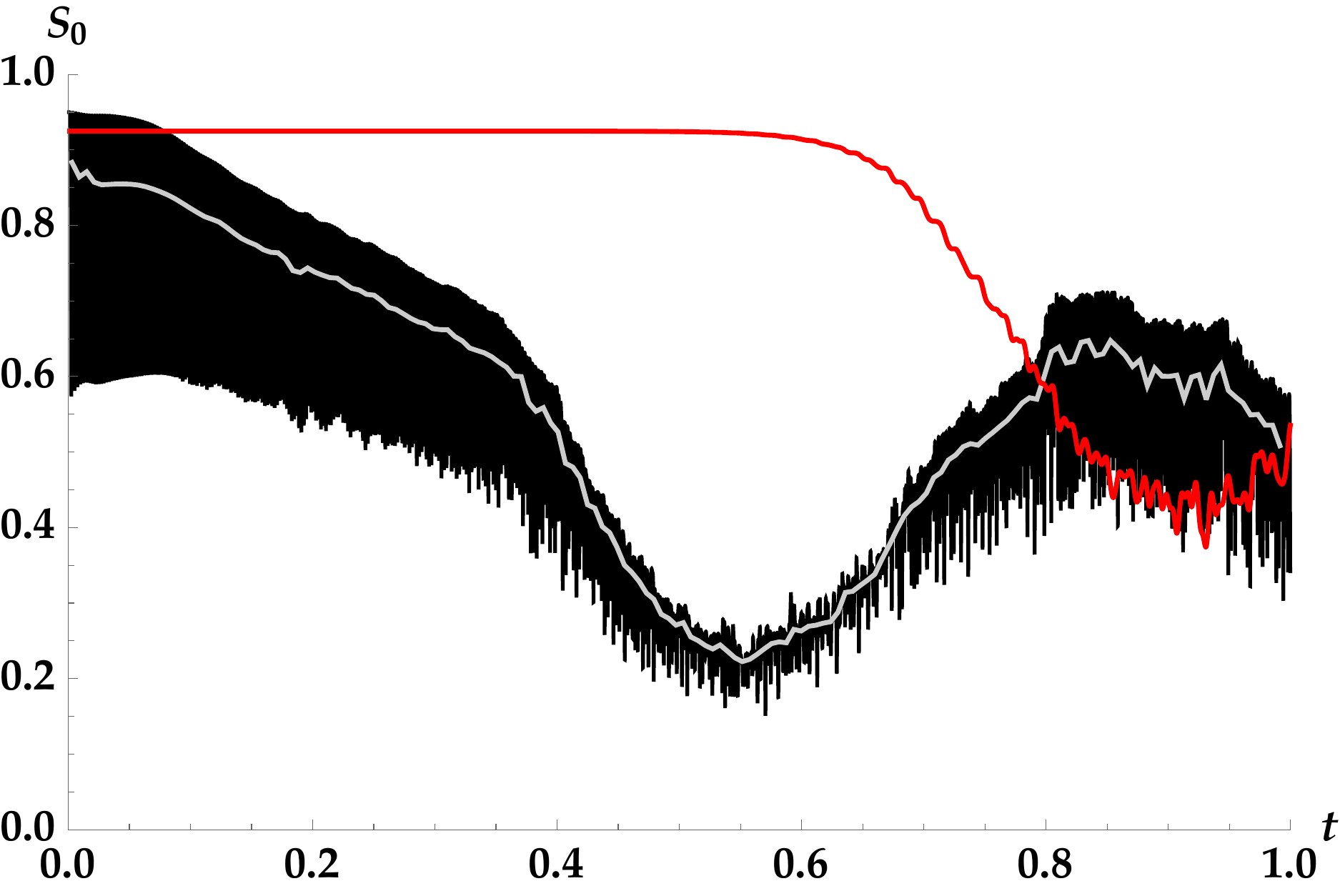}
}
\end{subfigure}
\begin{subfigure}{
\centering
\includegraphics[width=.43\textwidth]{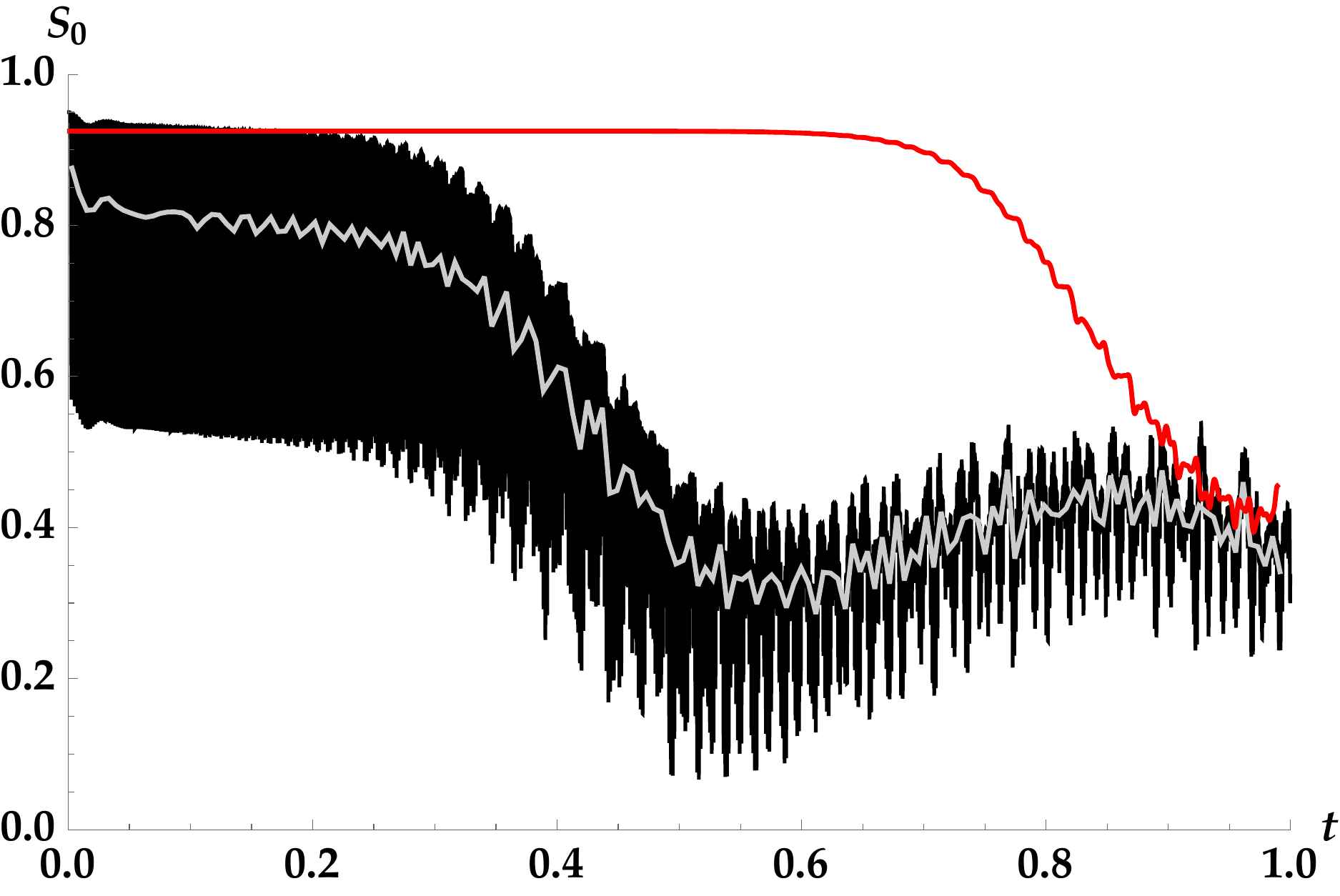}
}
\end{subfigure}
\caption{Same as the top four panels of Fig.~\ref{p0z_6000m}, but the calculations here were done using $l_\textrm{max} = 199$. Significant discrepancies, including spurious decoherence, set in at late times due to truncation of the angular-moment expansion.}
\label{p0z_200m}
\end{figure*}

\begin{figure*}
\centering
\begin{subfigure}{
\centering
\includegraphics[width=.31\textwidth]{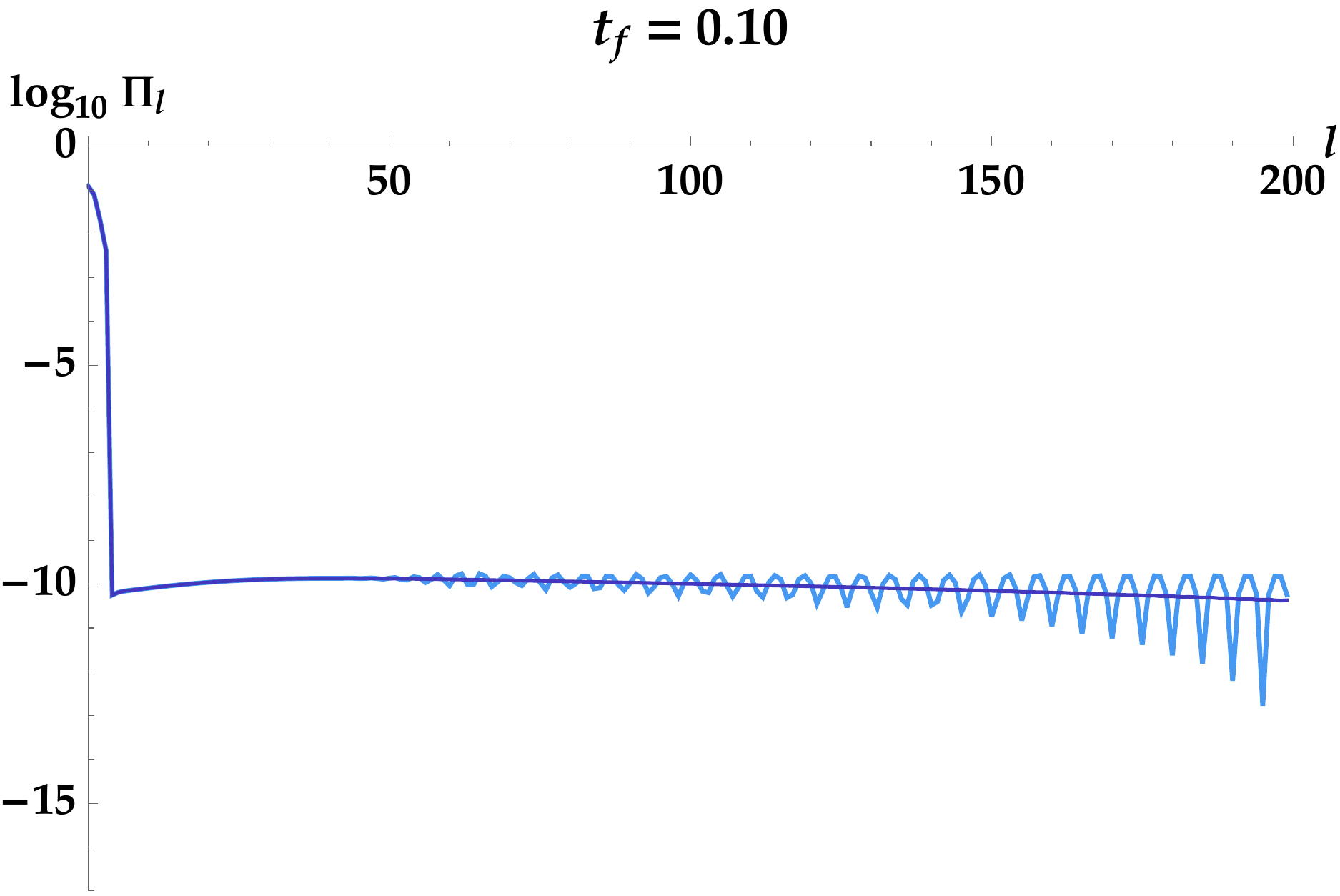}
}
\end{subfigure}
\begin{subfigure}{
\centering
\includegraphics[width=.31\textwidth]{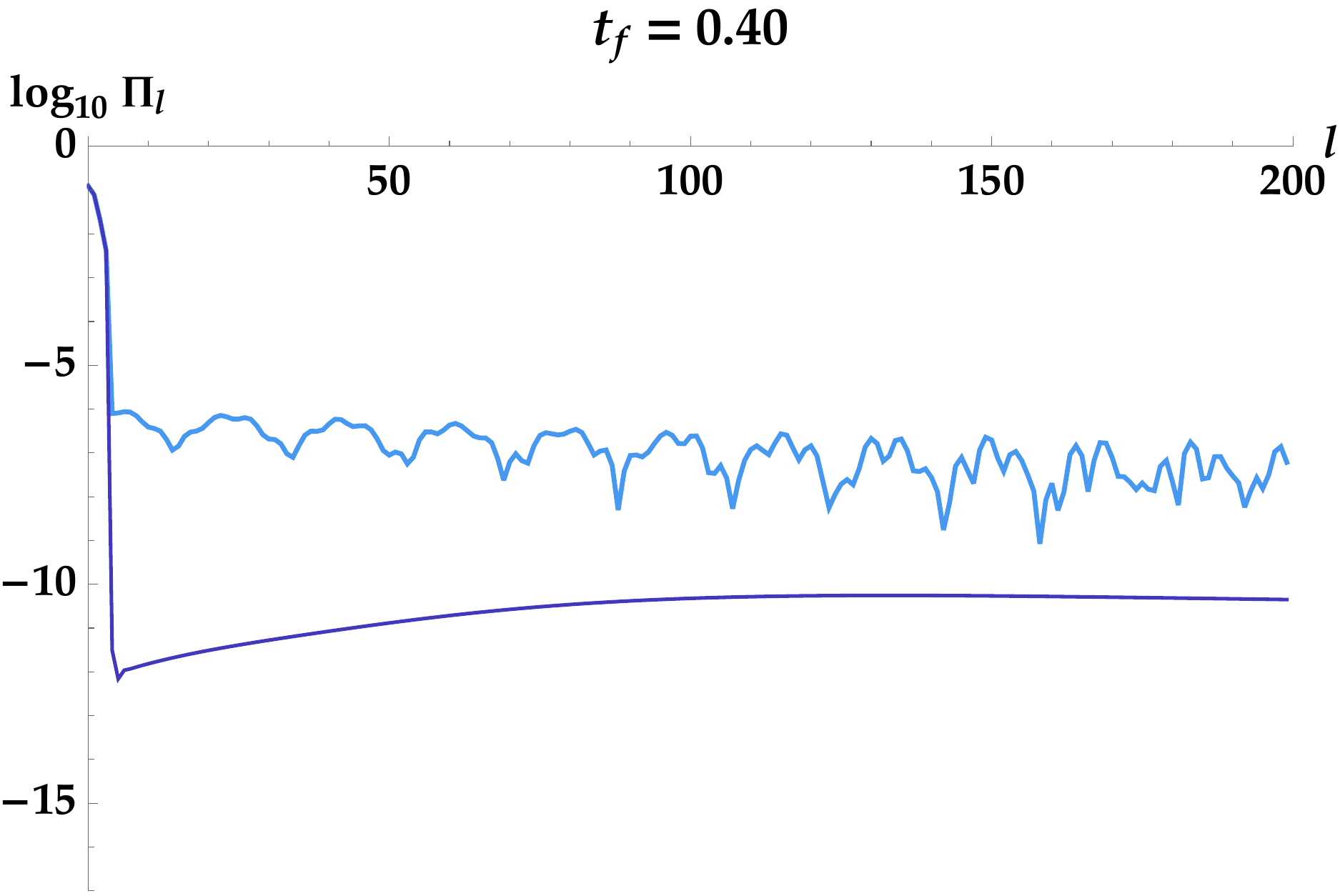}
}
\end{subfigure}
\begin{subfigure}{
\centering
\includegraphics[width=.31\textwidth]{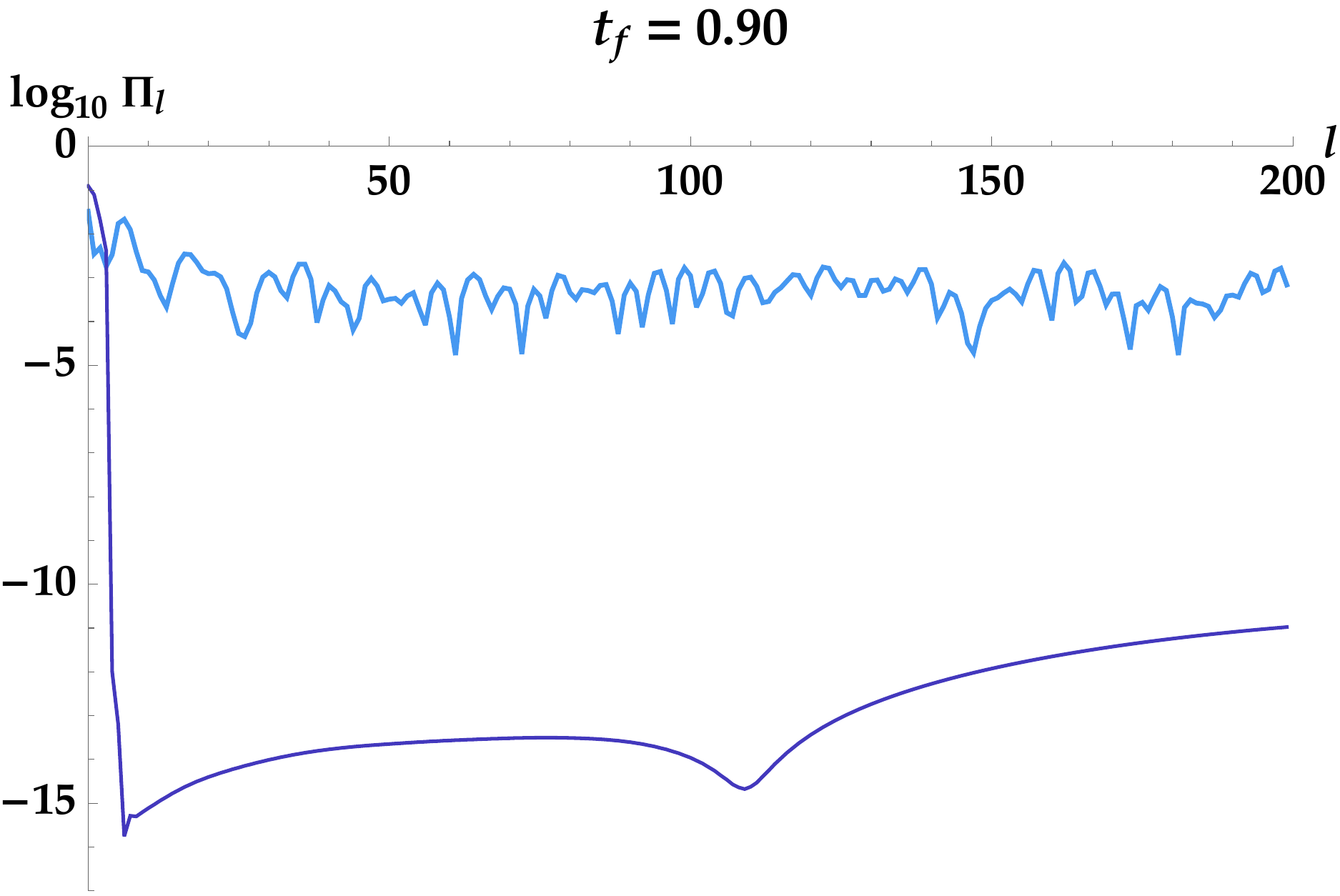}
}
\end{subfigure}
\caption{Angular power spectra, comparing a calculation (NH, $\alpha = 0.85$) with $l_\textrm{max} = 199$ (light blue) to one that is converged in $l_\textrm{max}$ (dark blue). In the former case, error accrues at $l_\textrm{max}$ and propagates back to lower $l$. Power artificially builds up until backreaction incites spurious decoherence.}
\label{pilvsl_200m_slow}
\end{figure*}

\begin{figure*}
\centering
\begin{subfigure}{
\centering
\includegraphics[width=.31\textwidth]{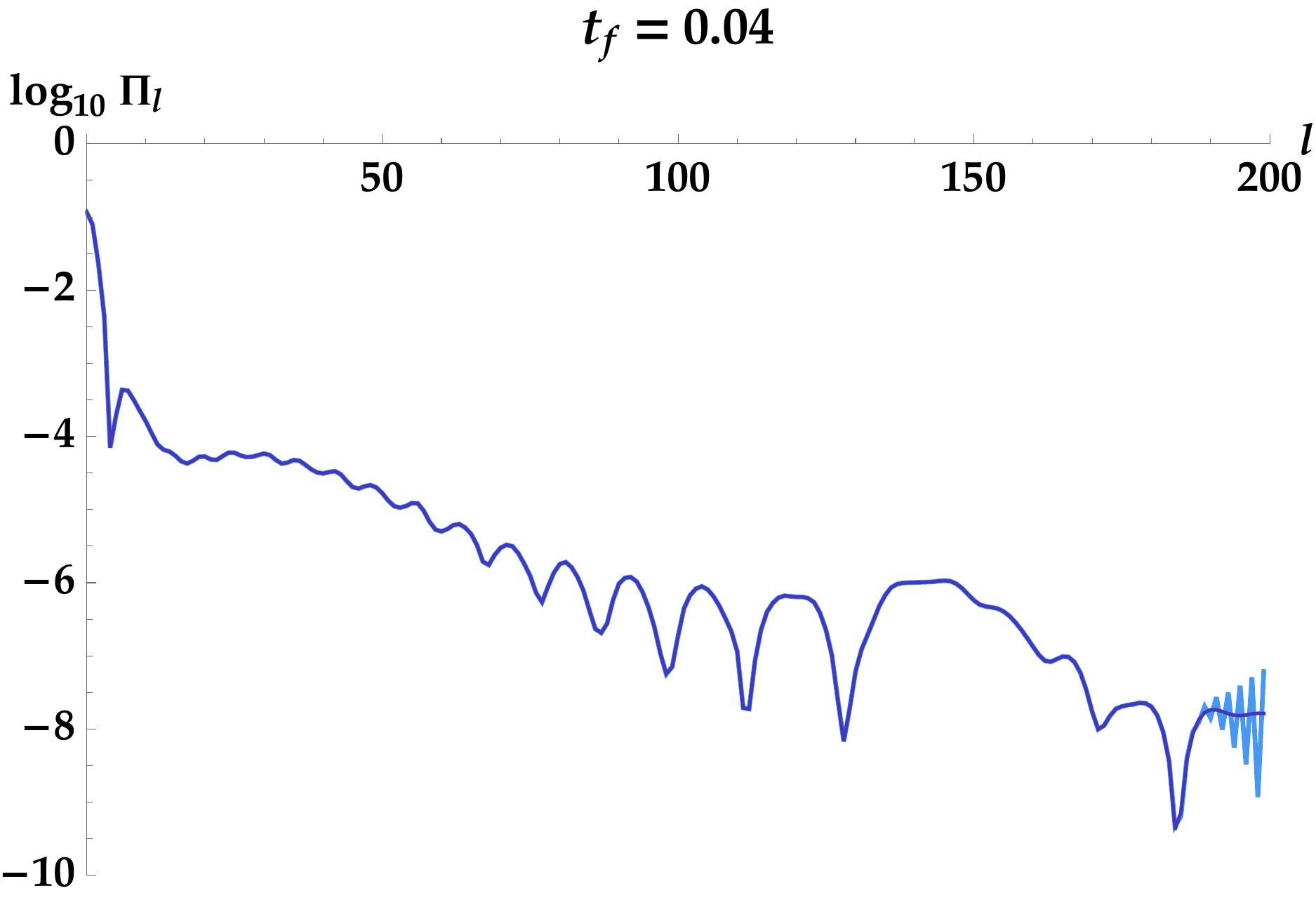}
}
\end{subfigure}
\begin{subfigure}{
\centering
\includegraphics[width=.31\textwidth]{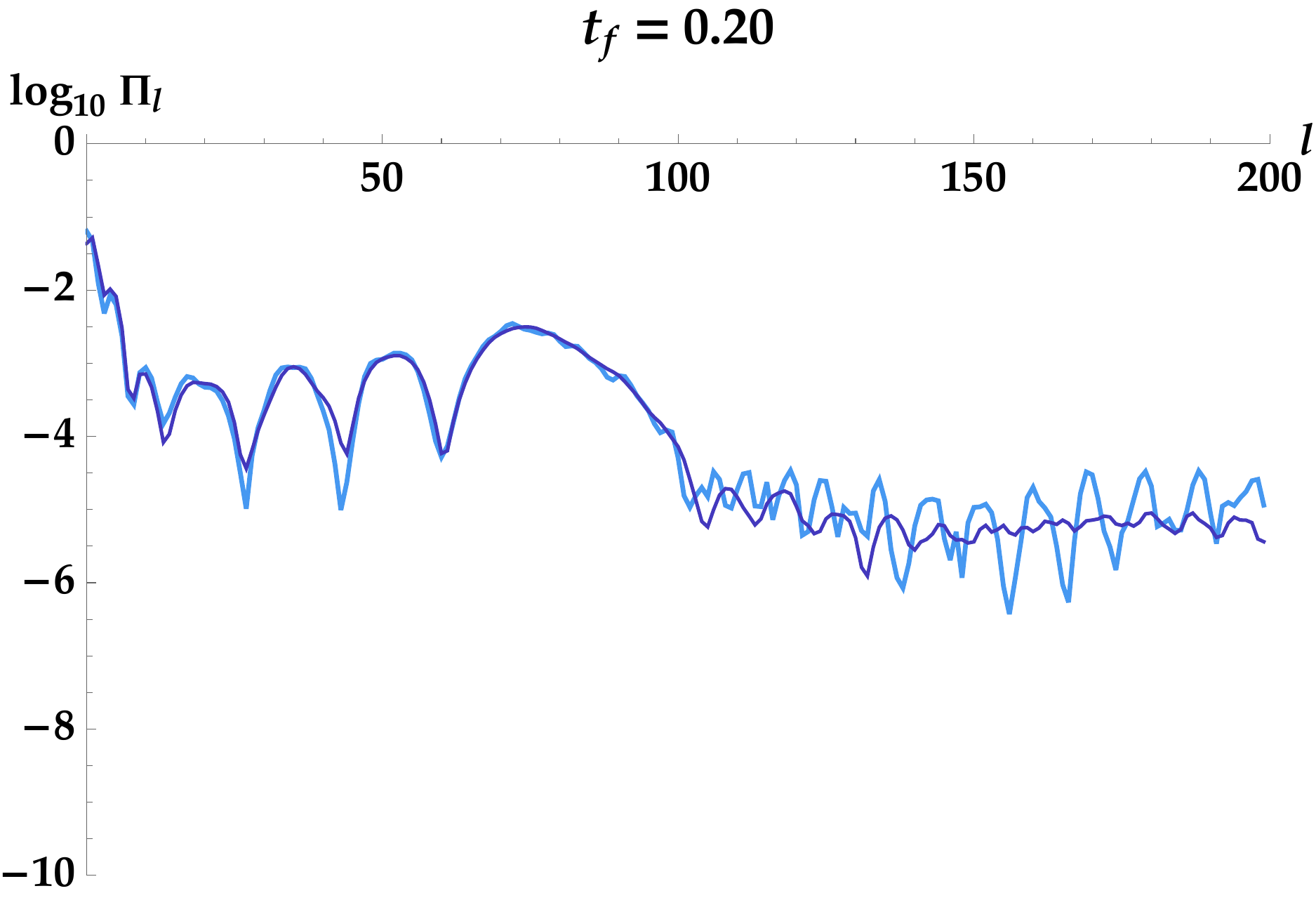}
}
\end{subfigure}
\begin{subfigure}{
\centering
\includegraphics[width=.31\textwidth]{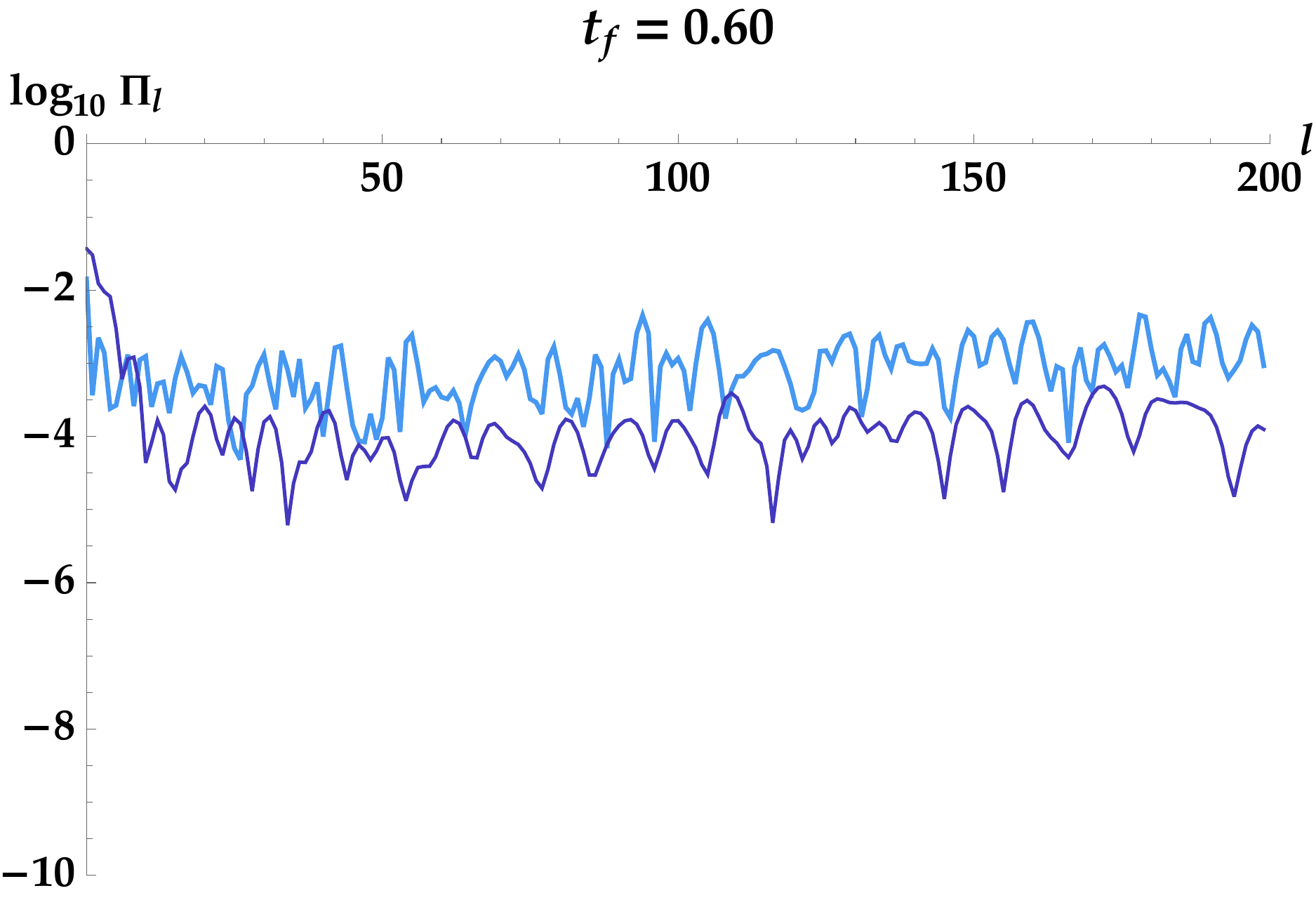}
}
\end{subfigure}
\caption{Same as Fig.~\ref{pilvsl_200m_slow}, but with $\alpha = 0.90$. Errors associated with truncation at $l_\textrm{max}$ again cause highly spurious evolution. Note that the sampled times and the range of the vertical axis differ with respect to the previous figure.}
\label{pilvsl_200m_fast}
\end{figure*}

\section{Numerical challenges \label{numerical}}

\begin{figure*}
\centering
\begin{subfigure}{
\centering
\includegraphics[width=.43\textwidth]{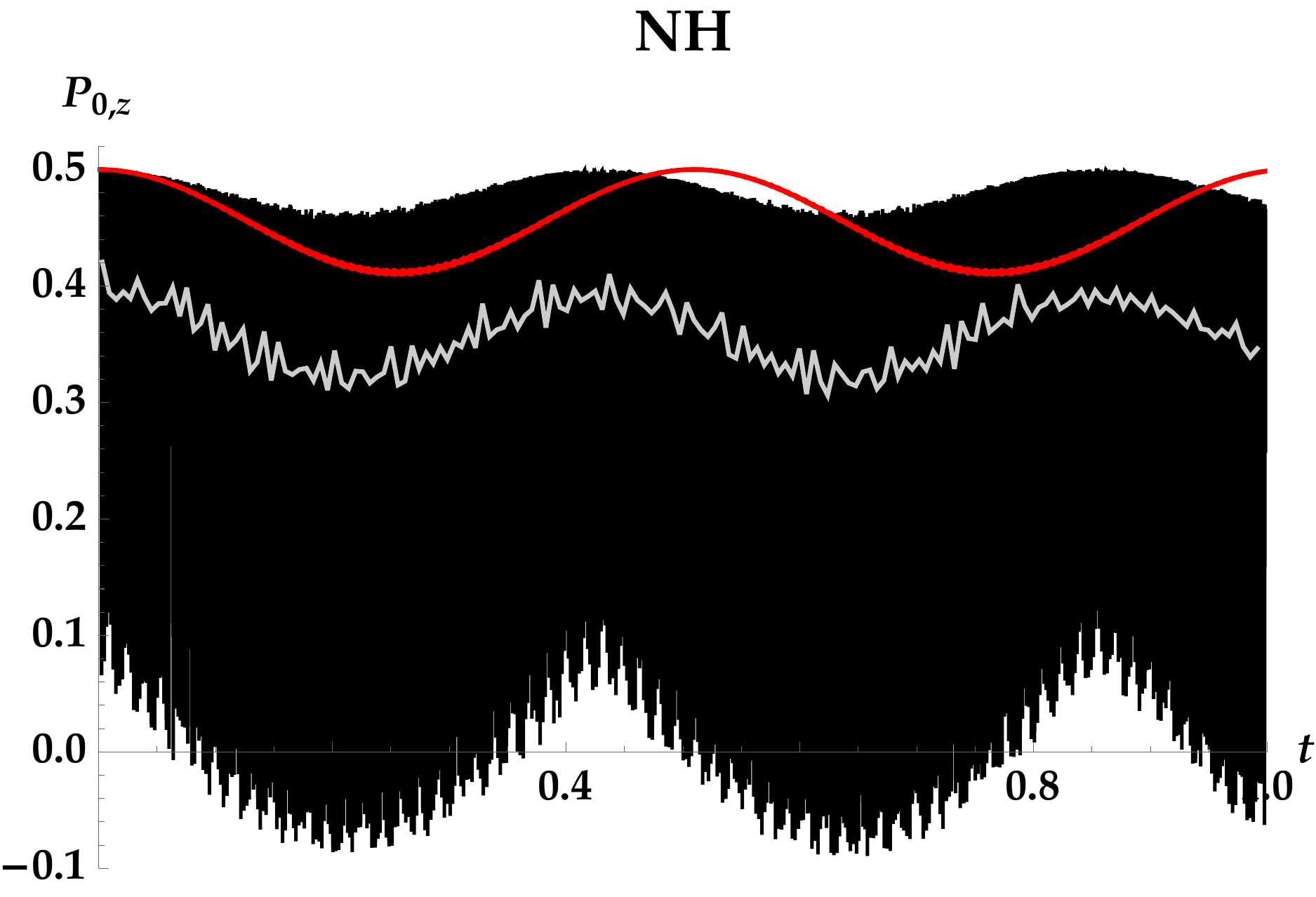}
}
\end{subfigure}
\begin{subfigure}{
\centering
\includegraphics[width=.43\textwidth]{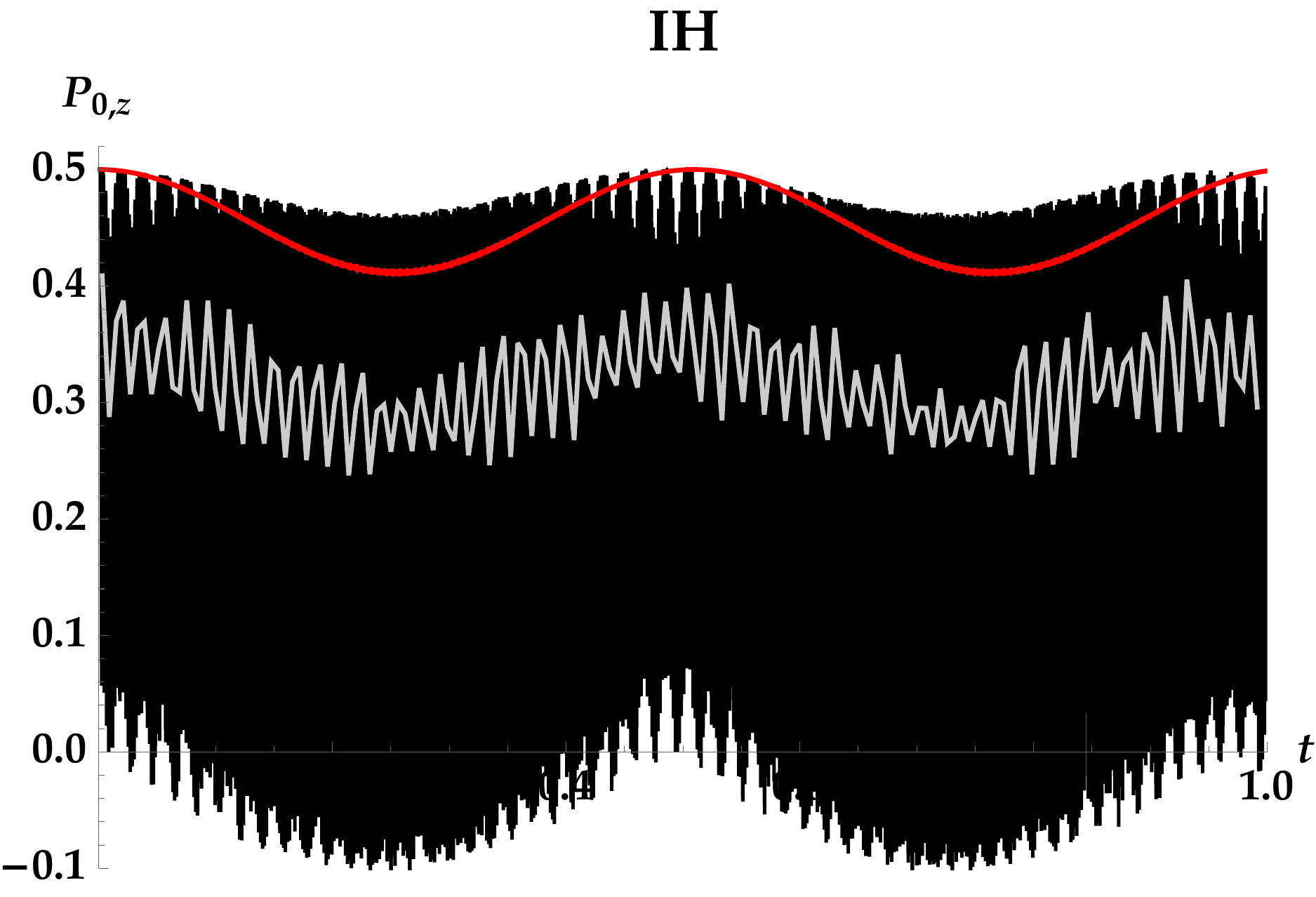}
}
\end{subfigure}

\begin{subfigure}{
\centering
\includegraphics[width=.43\textwidth]{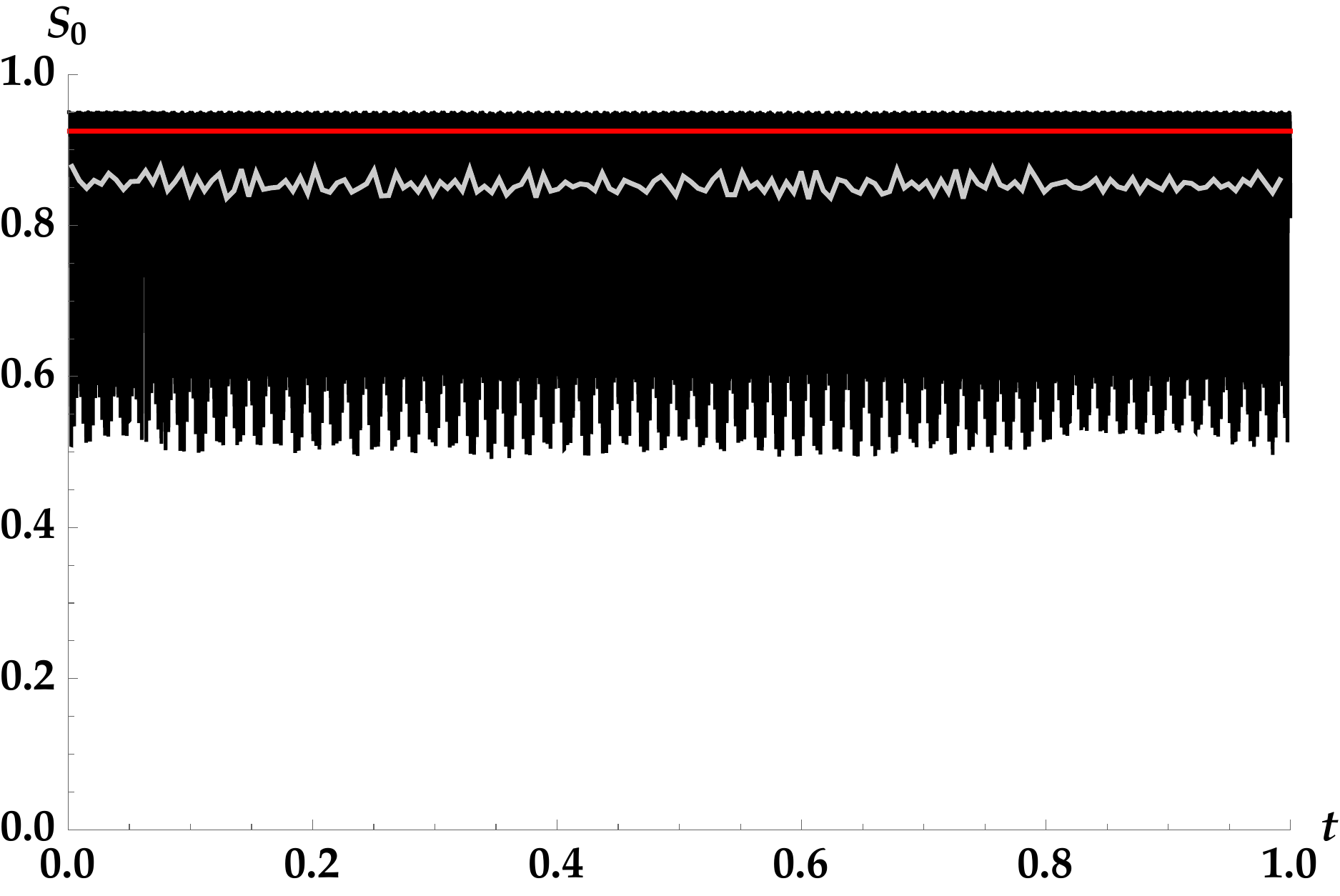}
}
\end{subfigure}
\begin{subfigure}{
\centering
\includegraphics[width=.43\textwidth]{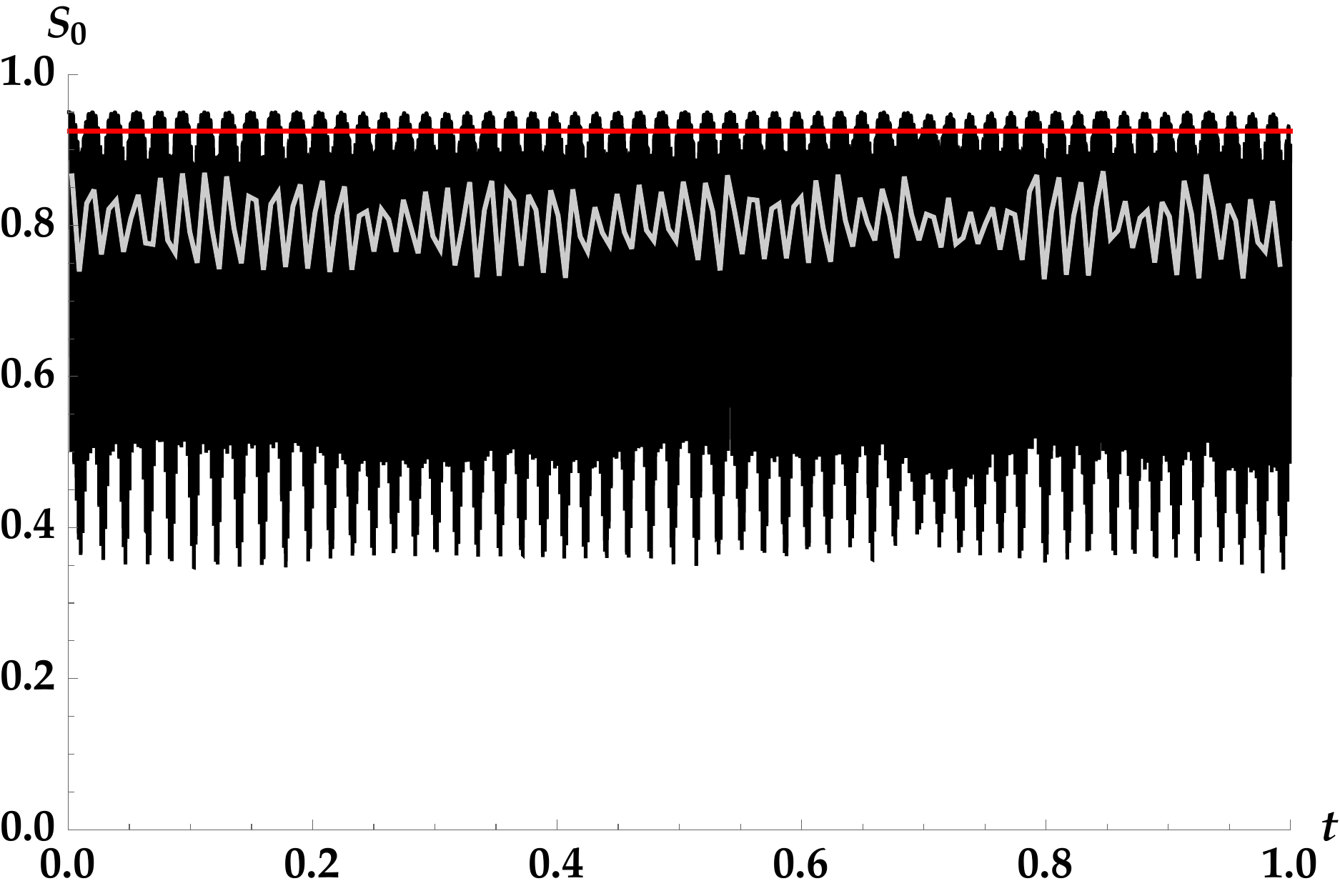}
}
\end{subfigure}
\caption{Same as the top four panels of Fig.~\ref{p0z_6000m}, but the calculations here were done using $l_\textrm{max} = 3$. Because $l \leq 3$ are the most relevant for evolution of the slow and fast pendulums, important features of the evolution are decently well approximated. Relaxation is \textit{not} captured, however, because momentum-space cascade is prohibited by the small number of angular moments.}
\label{p0z_4m}
\end{figure*}

In the previous section we observed the continuous transfer of power to smaller angular scales. In some instances, though, such as when the system is stable to FFC, the cascade is of such a modest magnitude that kinematic decoherence is negligible. It might be supposed that the cascade is therefore harmless and irrelevant. That conclusion, we now show, would be incorrect. 

Spurious evolution occurs when the angular-moment expansion is truncated at too small a value of $l$. Fig.~\ref{p0z_200m}, based on calculations with $200$ moments, exemplifies this point. Here all $\mathbf{P}_l$ and $\mathbf{\bar{P}}_l$ are set to zero at all times for $l \ge 200$. The figure can be directly compared to Fig.~\ref{p0z_6000m}, which is converged in the number of moments. Although the calculations agree at early times, discrepancy eventually sets in. The most dramatic difference is that the calculations with 200 moments are beset by spurious decoherence. Tellingly, its onset is largely independent of mass hierarchy but does depend on whether FFC is present or not.

The root of the problem is especially clear in the $\alpha = 0.85$ calculations. Fig.~\ref{pilvsl_200m_slow} shows the angular power spectrum $\log_{10} \Pi_l$, comparing the 200-moment calculations to the converged ones. Once the power at $l=199$ becomes nonzero, errors due to truncation start to accrue and propagate back to lower moments. As time proceeds, power artificially builds up in the ``inertial range'' rather than passing through $l=199$ out toward infinity. The build-up, through its backreaction on large angular scales, causes the spurious decoherence seen in the preceding figure. By the end of the 200-moment calculation, the time-averaged power is nearly equipartitioned over $0 \leq l \leq 199$.

The situation is similar in the $\alpha = 0.90$ calculations, as shown in Fig.~\ref{pilvsl_200m_fast}. Errors arise at $l=199$, propagate backward, and disrupt the low-$l$ evolution. Ultimately the system tends toward moment equipartition. The artificial build-up of power in the inertial range is less pronounced here, but still evident. Because FFC amplifies and quickens the cascade of power, spurious features arise earlier in the presence of FFC than in its absence.

It is well known that multi-angle calculations of neutrino flavor evolution are plagued by spurious instabilities. The traditional solution, working with a discretized approach, has been to chop up the angular coordinate very finely. In bulb-model computations, the total number of angle bins is typically $\mathcal{O}(10^3)$, if not more. Given that modern discrete-ordinate supernova simulations employ $\mathcal{O}(10)$ bins \cite{nagakura2018}, the apparent need to evolve 1000s of bins to track oscillations is clearly alarming.

Past work aimed at understanding the origin of spurious instabilities has demonstrated the appearance of erroneously unstable modes in the linear regime \cite{sarikas2012c, morinaga2018}. As the angular resolution is increased, the unstable modes proliferate in number but migrate toward the real axis of the complex plane of frequency $\Omega$ (for temporally growing modes) or wave number $K$ (for spatially growing ones). It has been argued that by working with integral quantities instead of discretizing, problematic logarithms are explicitly retained in the linear analysis and spurious instabilities are consequently avoided \cite{morinaga2018, johns2020}. (Possible evidence against this is found in Ref.~\cite{sarikas2012c}, whose authors reported that spurious instabilities still appeared in their analysis when using angular moments. Since they did not go into detail on this point, it is difficult for us to interpret with their finding.)

If this argument is correct, then spurious instabilities are more properly thought of as numerical artifacts of discretization, as opposed to resolution. The distinction is that spurious instabilities never fully disappear for discretized distributions, no matter how fine the resolution is, whereas they never appear for distributions expanded in basis functions, no matter how coarse the resolution is. This viewpoint is physically appealing, as there is no reason (that we are aware of) for thinking that collective instabilities are acutely sensitive to the details at extremely small scales. It also seems to be evidenced by Refs.~\cite{sarikas2012c} and \cite{morinaga2018}, which both found that physical modes are already rather well captured with very few bins. Rephrasing slightly, the problem is not that a low-resolution calculation is unaware of the real physics. The problem is that it also knows about the spurious side effects of discretization.

The spurious evolution we are addressing in this paper is due to errors in the evolution of $\mathbf{P}_{l_\textrm{max}}$ and $\mathbf{\bar{P}}_{l_\textrm{max}}$ being sequentially communicated to larger angular scales through the $\mathbf{D}_1$-mediated coupling of moment $l$ to moment $l - 1$. Even if angular moments evade the problem of spurious evolution as the term has usually been used (\textit{i.e.}, in reference to instabilities apparently arising from discretization), they do not escape this kind. We surmise that backreaction may have been responsible for the limitations of the multipole calculations reported on in Ref.~\cite{duan2014}, but again, lacking details, we cannot be sure.

Another closely related though distinct numerical phenomenon is the recurrence effect described in Ref.~\cite{raffelt2007b}. Recurrence occurs in the moment-truncated system because the number of degrees of freedom has been rendered artificially finite. Power is reflected at $l_\textrm{max}$ and, after a long enough period, the initial system is (nearly) restored to its initial configuration. As the authors of that paper note, nonlinearity is subdominant in their toy model demonstrating recurrence, allowing for the exponential solution to be effectively reversed as power returns to $l=0$. In our calculations, where kinematic decoherence is effected by the dephasing mechanism, power is scrambled over the finite range of moments as power reflected at $l_\textrm{max}$ interacts with power cascading from $l = 0$. We believe that the lack of recurrence in our calculations is, from another perspective, related to the different drift--diffusion behavior [Eq.~\eqref{drift}] that results from the relaxation process of interest to us here.

It is also worth distinguishing the build-up/backreaction problem from another numerical pitfall. Step size is always an issue in the numerical integration of differential equations, but in this system it is particularly critical that a small enough step size be used that $E_S$ is conserved to high precision. Failure to do so can enable the (often rather sudden) onset of spurious decoherence---a problem not quite the same as the step-by-step accumulation of error. We have found in our Runge--Kutta calculations that, for the same step size or error criterion, nonconservation of $E_S$ is typically much more severe when FFC occurs. The spurious onset of decoherence due to integration error is, once more, closely related to but distinct from the onset due to moment truncation.

In Ref.~\cite{johns2020} it was pointed out that the moments $l \leq 3$ are indispensable for homogeneous, axially symmetric FFC. The equation of motion of the fast pendulum, Eq.~\eqref{fastpendulum}, involves these four moments explicitly. Retaining only $l \leq 2$ makes FFC impossible, since the pendulum is never unstable if the gravity vector $\mathbf{G}'$ [Eq.~\eqref{gravity}], a rotated and scaled version of $\mathbf{D}_3$, always vanishes. Fortuitously, supernova simulations using M1 closure do provide the first four moments of the classical angular distributions. 

For these reasons it is interesting to look at the results of evolving only the first four moments. Fig.~\ref{p0z_4m} shows the same calculations, but now using $l_\textrm{max} = 3$, that were plotted in Fig.~\ref{p0z_6000m} (converged in $l_\textrm{max}$) and Fig.~\ref{p0z_200m} ($l_\textrm{max} = 199$). The results, including the fast oscillations, are not entirely dissimilar from those presented in Fig.~\ref{p0z_6000m}. The principal difference is that relaxation is virtually nonexistent in the calculations with only four moments, owing to the fact that there is no inertial range to facilitate kinematic decoherence. Not having an inertial range turns out to be helpful for capturing fast collective oscillations because power is unable to build up artificially and backreact on large scales, but the price is that the realistic transfer of power to small scales cannot occur either.

We suspect this finding might have some utility in more sophisticated approaches, for instance if the full flavor transformation can be approximated by taking the $l \leq 3$ evolution and modulating it with a prescription for relaxation. The guiding idea behind such a technique would be that, although relaxation is important, and cascade necessary for achieving it, the details at high $l$ are not critical. If an approach of this kind can be made to work, it would still remain to be seen whether it generalizes to more realistic models, where there may not be as neat a division between large-scale ``pendulum moments'' and small-scale ``cascade moments.''

As mentioned in Sec.~\ref{analogues}, the breakdown of numerical calculations due to increasingly fine phase-space structure is a familiar problem in plasma kinetics. Various strategies have been developed in response. Early work established that the breakdown is caused by approximating an infinite spectrum with a finite one and proposed the addition of small imaginary parts to the truncated set of eigenvalues, either by analytically extrapolating past the evolved modes or explicitly adding damping or collisional terms to the equations of motion \cite{grant1967, joyce1971}. Other approaches include periodically filtering the phase-space distribution in such a way as to smooth out the fine features \cite{cheng1976}, or allowing small-scale information to escape the calculation by imposing outgoing-wave boundary conditions \cite{eliasson2001}. All of these methods artificially generate entropy over time, but it is also possible to implement filtration in a manner that does not \cite{klimas1987}. More details and references on numerically solving the Vlasov equation can be found in Ref.~\cite{eliasson2010}.

These ideas---extrapolation, damping, filtration, boundary conditions---need to be adapted to neutrino quantum kinetics, and at this point we cannot vouch for their usefulness. We mention them here only as possible sources of inspiration for future efforts. Our own investigations have been limited, and on the whole we have found that the evolution is somewhat delicate to intervention. If strong artificial damping is applied to all $l \geq l_d$, for example, the reflection problem is not avoided. (As the damping rate is made \textit{very} large, the equations effectively resemble those of the system truncated at $l_d$.) Artificial damping that grows with $l$ has shown some promise, but our exploration of this approach has been far from exhaustive.

These plasma-inspired techniques are alternatives to the one described in connection to Fig.~\ref{p0z_4m}, in which only a small number of moments are evolved and relaxation is superimposed on the system in a physically motivated but non-self-consistent way. The plasma strategies are a compromise, evolving just enough of the cascade to capture relaxation at larger scales. At the other extreme is the approach traditionally taken in oscillation calculations: attaining convergence through very high resolution. We emphasize once more that relying on the last of these approaches continues to be a major impediment to progress in computing neutrino flavor evolution, and it may in fact be unnecessary.

\section{Discussion \label{conc}}

In this paper we have studied relaxation of the neutrino flavor field via momentum-space cascade, operating under the assumptions of monochromaticity, homogeneity, axial symmetry, and collisionless evolution. These simplifications are highly restrictive---axial symmetry and homogeneity above all, most likely---and preclude our calculations from being anything like \textit{predictions} of the flavor evolution taking place at specific times and locations in a supernova. Simplicity is what we are after, however. Our objective is not, at this point, to simulate flavor evolution, but rather to understand what the essential ingredients are and determine how they can be captured with both fidelity to the physics and deference to finite computational resources.

In calculations with parameters motivated by simulation data, we have seen that if kinematic decoherence does not occur through an exponential instability, it instead occurs through the comparatively gradual seepage of power to smaller angular scales. Fast oscillations expedite this process, sometimes considerably, by enhancing the dephasing of the relevant polarization vectors. We have situated this relaxation mechanism alongside the other major features of the model: bipolar oscillations, fast oscillations, and the multi-zenith-angle instability.

Cascade of power is a serious numerical concern, as errors at $l_\textrm{max}$ propagate back to larger scales, growing in magnitude until their backreaction on the isotropic moment is enough to spuriously decohere it. The computations performed for this study evolved $\mathbf{P}_l$ and $\mathbf{\bar{P}}_l$ directly. We expect similar spurious results to appear in calculations that evolve discretized angle bins, since the origin of the problem---the development of features at small angular scales---is a real aspect of the physics.

The problem is not unique to neutrino flavor, either. It is commonly encountered in weakly collisional kinetic systems of all kinds. We believe this is reason for optimism. Various strategies have already been developed in the context of kinetic plasmas for addressing essentially the same numerical challenges. If analogous techniques can be employed successfully for neutrino quantum kinetics, it may be possible to considerably lighten the computational burden of oscillation calculations. The key hypothesis we are advancing, which would make such computational strides possible, is that the evolution of the flavor field should not depend qualitatively on the details at very small angular scales. 

In reality, collisions damp the angular power spectrum, especially in the high-$l$ region. The scattering rate is typically much smaller than $\mu$, however, and collisions on their own are unlikely to resolve the numerical challenges of cascade. Speaking from a computational standpoint, we suspect that power still makes its way out to unacceptably large $l$. On the other hand, collisions have also been noted to exert a less direct but still possibly significant influence through the halo effect \cite{cherry2012, sarikas2012b, cirigliano2018}. From the angular-moment perspective, the result of the halo effect is to isotropize power initially spread over many moments, potentially giving that power (\textit{i.e.}, the scattered particles) greater leverage on the overall flavor evolution.

Having emphasized the numerical significance of our findings, let us return to the physical implications. Relaxing the assumption of axial symmetry permits additional instabilities, both fast and slow. At a more fundamental level, the dimensionality of phase space is simply larger. Whether or not the flavor field becomes unstable as a result of axial asymmetry, there are more channels for collisionless relaxation. There are, in other words, more ways to shift power from large angular scales to small ones. Crucially, the isotropic multipole can be decohered by any of the three vectors $\mathbf{D}_1^0$, $\mathbf{D}_1^{\pm1}$, where $\mathbf{D}_l^m$ is the $(l, m)$ spherical harmonic of the difference vector. Each of these three vectors interacts with the others, the effect of which is probably to disrupt any would-be periodic behavior even beyond the extent that we have seen here. This may already be apparent in the axially asymmetric calculations of Ref.~\cite{shalgar2020}.

With inhomogeneity comes a spectrum of instabilities associated with the nonvanishing convective term, and the dimensionality of phase-space is again expanded. Coordinate-space transfer differs in a number of significant ways from momentum-space transfer, and the manner in which our results are modified by convection is likely to depend on the nature of the inhomogeneity itself. What we can say, quite generally, is that we expect phase-space relaxation to figure prominently in realistic settings where the flavor field is unstable---with all of the ramifications that go along with it.

\begin{acknowledgments}
We are grateful to Evan Grohs, Kei Kotake, Zidu Lin, Amol Patwardhan, and Manibrata Sen for stimulating conversations at the INT workshop on ``Neutrinos from the Lab to the Cosmos'' and to Shashank Shalgar and Irene Tamborra for enjoyable correspondence. Support for this work was provided by NASA through the NASA Hubble Fellowship grant \# HST-HF2-51461.001-A awarded by the Space Telescope Science Institute, which is operated by the Association of Universities for Research in Astronomy, Incorporated, under NASA contract NAS5-26555. LJ and GMF acknowledge support from NSF Grant No. PHY-1914242, from the Department of Energy Scientific Discovery through Advanced Computing (SciDAC-4) grant register No. SN60152 (award number de-sc0018297), and from the NSF N3AS Hub Grant No. PHY-1630782 and Heising-Simons Foundation Grant No. 2017-22. AB acknowledges support from the U.S. Department of Energy Office of Science and the Office of Advanced Scientific Computing Research via the Scientific Discovery through Advanced Computing (SciDAC4) program and Grant DE-SC0018297 (subaward 00009650). In addition, he acknowledges support from the U.S. NSF under Grants AST-1714267 and PHY-1804048.
\end{acknowledgments}

\bibliography{all_papers}

\end{document}